\documentclass[twocolumn,showpacs,preprintnumbers,amsmath,amssymb]{revtex4-1}
\usepackage{graphicx}
\usepackage{dcolumn}
\usepackage{bm}
\usepackage{color}
\usepackage{enumitem,kantlipsum}
\usepackage{bm}
\usepackage[normalem]{ulem}
\usepackage{color}
\numberwithin{equation}{section}
\begin{document} 
\tighten
\newcommand {\be}{\begin{equation}}

\newcommand {\ee}{\end{equation}}

\newcommand {\gvt}{{\gamma v_2}}

\newcommand {\Dp}{{D_\perp(\bq)}}

\newcommand {\Dpr}{{D_{\rho\perp}(\bq)}}

\newcommand {\xpa}{the direction of mean flock motion $\hat{x}_{_\parallel}$}

\newcommand {\bea}{\begin{eqnarray}}

\newcommand{\cred}{\color{red}}
\newcommand{\cmag}{\color{magenta}}
\newcommand{\cblue}{\color{blue}}

\newcommand{\cO}{{\cal O}}
\newcommand {\eea}{\end{eqnarray}}

\newcommand {\tq}{\theta_{{\bf q}}}

\newcommand {\bq}{{\bf q}}

\newcommand {\bqp}{{\bf q}_\perp}

\newcommand {\xpl}{{\hat{\bf x}_{_\parallel}}}

\newcommand{\vt}{{\bf v}_{_T}}

\newcommand {\vq}{\overline{|{\bf v}_{_\perp}({\bf q})|^2}}

\newcommand {\rsv}{\overline{|{\bf v}_{_\perp}({\bf r})|^2}}

\newcommand {\red}{\texttt{\color{red}}}
\newcommand {\blue}{\texttt{\color{blue}}}
\newcommand{\vp}{{\bf v}_{_\perp}}
\newcommand{\bv}{{\bf v}}
\newcommand{\qp}{{\bf q}_{_\perp}}
\newcommand{\brp}{{\bf r}_{_\perp}}
\newcommand{\rp}{ r_{_\perp}}
\newcommand{\chip}{ \xi_{_\perp}}
\newcommand{\chipar}{ \xi_{_\parallel}}
\newcommand{\rpar}{ r_{_\parallel}}
\newcommand{\br}{{\bf r}}
\newcommand{\bR}{{\bf R}}

\newcommand{\qpar}{ q_{_\parallel}}
\newcommand{\mqp}{ q_{_\perp}}

\newcommand {\fl}{\cite{TT1, TT2,TT3,TT4, NL}}

\def\lsim{\:\raisebox{-0.5ex}{$\stackrel{\textstyle<}{\sim}$}\:}

\def\gsim{\:\raisebox{-0.5ex}{$\stackrel{\textstyle>}{\sim}$}\:}

\def\Dtens{\mbox{\sffamily\bfseries D}}

\def\Wtens{\mbox{\sffamily\bfseries W}}

\def\Ptens{\mbox{\sffamily\bfseries P}}

\def\Otens{\mbox{\sffamily\bfseries O}}

\def\TT{\cite{TT1,TT2,TT3,TT4,NL}}

\def\Qtens{\mbox{\sffamily\bfseries Q}}

\def\Q{\mbox{\sffamily\bfseries Q}}

\def\Ntens{\mbox{\sffamily\bfseries N}}

\def\Ctens{\mbox{\sffamily\bfseries C}}

\def\Itens{\mbox{\sffamily\bfseries I}}

\def\Atens{\mbox{\sffamily\bfseries A}}

\def\A{\mbox{\sffamily\bfseries A}}

\def\Ktens{\mbox{\sffamily\bfseries K}}

\def\Vtens{\mbox{\sffamily\bfseries V}}

\def\Gtens{\mbox{\sffamily\bfseries G}}

\def\ftens{\mbox{\sffamily\bfseries f}}

\def\vtens{\mbox{\sffamily\bfseries v}}

\def\nabbold{\mbox{\boldmath $\nabla$\unboldmath}}

\def\nabvec{\mbox{\boldmath $\nabla$}}

\def\sigtens{\mbox{\boldmath $\sigma$\unboldmath}}

\def\etatens{\mbox{\boldmath $\eta$\unboldmath}}

\def\beq{\begin{equation}}

\def\bea{\begin{eqnarray}}

\def\eeq{\end{equation}}

\def\eea{\end{eqnarray}}


\title{Hydrodynamic Theory of Flocking in the Presence of  Quenched Disorder}
\author{John Toner}
\author{Nicholas Guttenberg}


\affiliation{Institute for Theoretical Science and  Department of Physics, University of Oregon, Eugene, OR 97403}
\author{Yuhai Tu}
\affiliation{IBM T. J. Watson Research Center, Yorktown Heights, NY 10598}


\begin{abstract}

The effect  of  quenched (frozen) orientational disorder on the collective motion of active particles is analyzed. We find that, as with annealed disorder (Langevin noise), active polar systems are far more robust against quenched disorder than their equilibrium counterparts. In particular, long ranged order (i.e., the existence of a non-zero average velocity $\langle {\bf v} \rangle$) persists in the presence of quenched disorder even in spatial dimensions $d=3$, while it is destroyed even by arbitrarily weak disorder
in $d \le 4$ in equilibrium systems. Furthermore, in $d=2$, quasi-long-ranged order (i.e., spatial velocity correlations that decay as a power law with distance) occurs when quenched disorder is present, in contrast to the short-ranged order that is all that can survive in equilibrium.  These predictions are borne out by simulations in both two and three dimensions.

 \end{abstract} 


\maketitle

\section{Introduction}\label{introduciton}

A great deal of the immense current interest  in ``Active Matter" focuses on  coherent collective motion, i.e.,  ``flocking" \cite{boids, Vicsek, TT1,TT2,TT3,TT4,NL}, also sometimes called ``swarming" \cite{dictyo,rappel1}, or by a variety of other names.
Such coherent motion occurs over enormous  numbers of
self-propelled entities,  and a wide range 
 of length scales:  from
kilometers (herds of wildebeest) to microns (microorganisms
Dictyostelium discoideum \cite{dictyo,rappel1}); to the submicron (e.g., mobile  macromolecules in living cells \cite{actin,microtub}).

Vicsek et. al. \cite{Vicsek} were the first to  note both the analogy between such coherent motion and ferromagnetic ordering, and its breakdown
in that coherent motion is possible even in $d=2$. This  apparent violation of  the Mermin-Wagner theorem \cite{MW} has been explained by the ``hydrodynamic" theory of flocking \cite{TT1,TT2,TT3,TT4,NL}, which shows that, unlike equilibrium ``pointers", non-equilibrium ``movers" {\it can}
spontaneously break a continuous symmetry (rotation invariance) by developing long-ranged orientational order  (as they must  to have a  non-zero average velocity $\left<{\bf v} ({\bf r}, t) \right>\ne \bf 0$), even in   noisy systems with  only short ranged interactions in spatial dimension $d=2$.

The mechanism for this apparent violation of the ``Mermin-Wagner" theorem \cite{MW} is fundamentally nonlinear   and non-equilibrium \cite{TT1,TT2,TT3,TT4,NL}.  Certain nonlinear terms in the hydrodynamic equations of motion become ``relevant", in the renormalization group (RG) sense, as the spatial dimension $d$ is lowered below 4, leading to a breakdown of linearized hydrodynamics \cite{FNS} which suppresses fluctuations enough  to stabilize long-ranged order  in $d=2$. 

In equilibrium systems, 
even {\it arbitrarily weak} random fields  destroy long-ranged ferromagnetic order in all spatial dimensions $d\le4$ \cite{Harris, Geoff, Aharonyrandom,Dfisher}. This
raises the question:
can the non-linear, non-equilibrium effects that make long-ranged order possible in 2d flocks without quenched disorder even stabilize them   when random field  disorder is present? This  issue was first  investigated by Chepizhko et.al. \cite{Peruani}, who simulated a model which,  though very different in its microscopic details, should be in the same universality class as the one  we consider here. More recently, Das et.al \cite{Das} have studied this problem both analytically (in a linearized approximation), and numerically in two dimensions, and also find quasi-long-ranged order in $d=2$.

In this paper, we address this problem  analytically, including non-linear effects, in both two and three dimensions, using the hydrodynamic theory of flocking developed in \cite{TT1,TT2,TT3,TT4,NL}, and through simulations. We consider only "dry" flocks; that is, flocks with no momentum conservation. We restrict ourselves to flocks in which the number of flockers is conserved; ``Malthusian" flocks \cite{Malthus}, in which the flockers are continuously being born and dying as the motion goes on, will be treated elsewhere \cite{future}. 

Both approaches confirm  that   flocks with non-zero quenched disorder {\it are}, indeed, far better ordered than their equilibrium analogs, i.e.,  ferromagnets subject to quenched random fields. Specifically, we find that flocks {\it can} develop long ranged order in three dimensions, and quasi-long-ranged order (defined below)  in two dimensions, in strong contrast to the equilibrium case, in which only short-ranged order is possible in both  three and two  dimensions \cite{Harris, Geoff, Aharonyrandom,Dfisher}.

By long-ranged orientational order, we mean  a non-zero average velocity $\overline{{\bf v} ({\bf r}, t)}\ne 
{\bf 0}$, where the overbar denotes an average over the quenched disorder. Since we believe that the 
often made "self-averaging" assumption (that is, that spatial averages calculated in a sufficiently large 
system for a {\it particular} generic realization of the quenched disorder 
will be equal to ensemble averages over the disorder ) applies to these flocks, our prediction that three-
dimensional flocks have long-ranged order in this sense implies that  a single large three dimensional 
flock in the presence of quenched disorder {\it can} have a non-zero spatially averaged velocity (the overbar and $\langle .. \rangle$ will be used interchangeably in this paper).  

 By quasi-long-ranged order, we mean the average velocity of a large flock is zero, but velocity correlations decay very slowly (specifically, algebraically) with distance:

\begin{equation}
\overline{{\bf v}({\bf r},t)\cdot{\bf v}({\bf r}^{~\prime},t)}\propto|{\bf r}-{\bf r}^{~\prime}|^{-\sigma(\Delta)}~,
\label{qlro}
\end{equation}
where the exponent $\sigma(\Delta)$ is non-universal  (that is, system dependent); specifically, it depends on the degree of quenched disorder, which is characterized in our model by a single parameter $\Delta$ (defined more precisely below).

Our prediction that quasi-long-ranged order occurs in two-dimensional flocks with quenched random field 
disorder agress   with the simulation results of Chepizhko et. al. \cite{Peruani}, and Das et.al. \cite{Das}.

We also find that the behavior of the propagating ``longitudinal sound modes" (that is, coupled density and velocity modes) in flocks radically affects the response of the flock to quenched disorder.
It has long been known \cite{TT1,TT2,TT3,TT4,NL} that
the speeds of these sound modes  are strongly 
anisotropic. Depending on the values of certain phenomenological parameters characterizing a flock - in 
particular, the speeds $\gamma$ and $v_2$ of the pure velocity and pure density modes for propagation 
in the direction of flock motion - this anisotropy can exhibit two qualitatively very different structures. 
When the product $\gamma v_2 > 0$, the speed of one of the sound modes vanishes when the angle $
\theta$ between the direction of propagation of the sound and the direction of mean flock motion satisfies 
$\theta=\pm\theta_c=\pm\arctan \left[{\sqrt{\gamma v_2}\over c_0}\right]$. (Here $c_0>0$ is the speed of sound propagating {\it perpendicular} to the direction of flock motion). As is obvious from this expression, when the 
product $\gamma v_2 < 0$, there is no $\theta_c$, and the speed  of the sound modes never vanishes for any direction of propagation.

We find that this difference between the cases $\gamma v_2 > 0$ and $\gamma v_2 < 0$ leads to radical 
differences in   the scaling behavior of these systems. The case $\gamma v_2 < 0$ proves to be 
completely analytically tractable; we can determine exactly, without approximations or assumptions, the 
scaling laws governing the long-distance and long-time behavior of the flock. In two dimensions, for this case, we can argue compellingly for the existence of quasi-long-ranged order (Eq.~(\ref{qlro})). Furthermore, in three dimensions, we find exact scaling laws for the  velocity fluctuations. For example, the connected two point velocity correlation function obeys:

\begin{eqnarray}
C_{vv}({\bf r})&\equiv&\overline{\delta{\bf v}({\bf r}+{\bf R},t)\cdot\delta{\bf v}({\bf R},t)}\nonumber\\&=& r^{-{1\over 2}} f_T\left(\left({r\over\xi}\right)^{1\over 4 } \sin\theta_r 
\right)\nonumber\\&\propto&\left\{
\begin{array}{ll}
\left(\sin\theta_r\right)^{-{2\over 3 }} r^{-{2\over 3 }}\,\,\,\,\,\,\,\,\,,
&{\theta_r}\gg \left({r\over\xi}\right)^{-{1\over 4 }}\,,
\\ \\
r^{-{1\over 2}}\,\,\,\,\,\,\,\,\,,
&{\theta_r}\ll \left({r\over\xi}\right)^{-{1\over 4 }}\,,
\end{array}\right.
\label{Creal1}
\end{eqnarray}
where $\delta{\bf v}({\bf r}^{~\prime},t)\equiv{\bf v}({\bf r},t)-\overline{{\bf v}}$, $\theta_r$ is the angle between ${\bf r}$ and the direction of propagation, $\xi$ is a characteristic length that depends on the flock, and strongly on the strength $\Delta$ of the disorder, and 
the exponents $2/3$, $1/2$, and $1/4$ are {\it exact}. 

Note that the exponents in Eq.~(\ref{Creal1}) are {\it not} those predicted by a linearized version of our theory; the non-linearities change these exponents substantially. In fact, the purely linearized theory predicts that there is no long-ranged-order at all in $d=3$; that is, that $\overline{{\bf v}}={\bf 0}$, always, in $d=3$.  The full, nonlinear theory shows that this is not the case, and that $\overline{{\bf v}}\ne{\bf 0}$ for sufficiently small, but non-zero, disorder strength $\Delta$.

In the case $\gamma v_2 > 0$, the situation is less clear. While we can show in this case that non-linearities {\it do} make the behavior of the flock different from that predicted by the linearized theory, and in particular that long ranged order  ($\overline{{\bf v}}\ne{\bf 0}$) survives in $d=3$, we cannot convincingly show that quasi-long-ranged order occurs in $d=2$. Nor can we obtain exact exponents in three dimensions. If we {\it assume}, however, that the ``convective" non-linearity is the dominant non-linearity in the flock dynamics, then we {\it can} demonstrate the existence of quasi-long-ranged order in $d=2$. We also thereby obtain predictions for  correlation functions in Fourier space which agree quantitatively with our simulations. Furthermore, there is considerable numerical and experimental evidence \cite{TT2, Rao, field} that this assumption that the convective non-linearity
dominates is correct in flocks with annealed disorder, which supports (although by no means proves) the correctness of this conjecture for the quenched disorder case.

We can still 
 predict scaling laws  for the velocity correlations in three dimensions even in the case $\gamma v_2>0$.

For example, the connected velocity autocorrelation function defined above  in $d=3$ 
is given by

\begin{eqnarray}
C_{vv}({\bf r})=C_L({\bf r})+C_T({\bf r})\,,
\end{eqnarray}
where $C_L({\bf r})$ and $C_T({\bf r})$   represent the contributions to $C_{vv}({\bf r})$ coming from ``longitudinal" (i.e., compressive) and ``transverse" (i.e., shear) fluctuations, 
and respectively
obey the scaling laws 

\begin{eqnarray}
C_L({\bf r})&=&r^{-{\Omega}} f_L\left(\delta\theta_r r^{\beta } 
\right) h_L(\theta_r)\nonumber\\&\propto&\left\{
\begin{array}{ll}
(\delta\theta_r r)^{2\chi}\,\,\,\,\,\,\,\,\,,
&{\delta\theta_r}\gg r^{-{\beta }}\,,
\\ \\
r^{-{\Omega}}\,\,\,\,\,\,\,\,\,,
&{\delta\theta_r}\ll r^{-\beta}\,,
\end{array}\right.
\label{CLreal}
\end{eqnarray}
and
\begin{eqnarray}
C_T({\bf r})&=&r^{-\Omega} f_T\left(\theta_r r^{\beta } 
\right) \nonumber\\&\propto&\left\{
\begin{array}{ll}
(\theta_r r)^{2\chi}\,\,\,\,\,\,\,\,\,,
&{\theta_r}\gg r^{-\beta}\,,
\\  \\
r^{-\Omega}\,\,\,\,\,\,\,\,\,,
&{\theta_r}\ll r^{-\beta}\,.
\end{array}\right.
\label{CTrealgen}
\end{eqnarray}

In Eq.~(\ref{CLreal}),  we've defined
$\delta\theta_r\equiv\theta_r+\theta_c-{\pi\over2}$, the function $h_L(\theta_r)$ is a smooth, analytic, $\cO(1)$ function of $\theta_r$, with no strong dependence on $\theta_r$ near $\theta=\pi/2-\theta_c$. The exponents $\beta$ and $\Omega$ in Eq.~(\ref{CLreal}) and Eq.~(\ref{CTrealgen}) are determined by the other two unknown, but universal, exponents: the anisotropy exponent $\zeta$, and the roughness exponent $\chi$, via the relations
\begin{equation}
\beta=1-{1\over\zeta}\,\,,\,\,
\Omega=-2{\chi\over\zeta}\,.
\end{equation}

Note that $C_L({\bf r})$ and the density correlation $C_{\rho\rho}({\bf r})$ (see Eq.~(\ref{Crhoreal})) exhibit their strongest anisotropies in different directions from those in which $C_T({\bf r})$ does: $C_L({\bf r})$ and $C_{\rho\rho}({\bf r})$ are most strongly anisotropic near $\theta_r={\pi\over2}-\theta_c$, while  $C_T({\bf r})$ is most strongly anisotropic near $\theta_r=0$. Thus, the full correlation function $C_{vv}({\bf r})$ exhibits strong anisotropy near both directions of ${\bf r}$.

While we can say nothing definite in $d=3$ for the case 
$\gamma v_2>0$, it is tempting to 
conjecture that the exponents  $\zeta$ and $\chi$ take on the same values as for $\gamma v_2<0$ in $d=3$, which are $\zeta=4/3$, $\chi=-1/3$. If this is the case, then we obtain $\beta=1/4$ and $\Omega=1/2$. We really have no justification for this conjecture, however, other than the fact that an analogous conjecture for flocks with {\it annealed} disorder appears empirically to get the correct exponents for $d=3$.

In all four cases,
density fluctuations exhibit long-ranged correlations, which also obey a simple scaling law:
\begin{eqnarray}
C_{\rho\rho}({\bf r})&\equiv&\overline{\delta{\rho}({\bf r}+{\bf R},t)\cdot\delta{\rho}({\bf R},t)}\nonumber\\ &=&r^{-{\Omega_\rho}} f_L\left(\delta\theta_r r^{\beta_\rho } 
\right) h_\rho(\theta_r)\nonumber\\&\propto&\left\{
\begin{array}{ll}
(\delta\theta_r r)^{2\chi_\rho}\,\,\,\,\,\,\,\,\,,
&{\delta\theta_r}\gg r^{-{\beta_\rho }}\,,
\\ \\
r^{-{\Omega_\rho}}\,\,\,\,\,\,\,\,\,,
&{\delta\theta_r}\ll r^{-\beta}\,,
\end{array}\right.
\label{Crhoreal}
\end{eqnarray}
which only shows strong anisotropy near $\theta_r={\pi\over2}-\theta_c$, the function $h_{\rho}(\theta_r)$ is a smooth, analytic, $\cO(1)$ function of $\theta_r$. 

In three of the four cases, namely, $d=2$, for both signs of $\gvt$, and $d=3$, $\gvt<0$ : $\beta_\rho=0$,  $\chi_\rho=d-2$, and $\Omega_\rho=2\chi_\rho=2(d-2)$. Another way to say this is that in these cases, $C_{\rho\rho}({\bf r})\propto r^{2-d}\times f_L(\theta_r)$; that is, $C_{\rho\rho}({\bf r})$ is proportional to $r^{2-d}$ for all directions $\theta_r$ of $\br$.  For $d=3$, $\gvt>0$: $\beta_\rho=\beta=1-{1\over\zeta}$,  $\chi_\rho=\chi$, and $\Omega_\rho=\Omega=-2{\chi\over\zeta}$.

As in flocks with annealed disorder, these long-ranged correlations lead to "Giant number fluctuations", as predicted and seen in both active nematics \cite{actnem} and flocks with annealed disorder \cite{GNF}: that is, if one counts the number of particles $N$ in a hypercubic subvolume, and look at  the mean squared fluctuations $\overline{ \Delta N^2 }$ scale with the mean number $\overline{N}$, we find that these fluctuations are much larger than the usual ``law of large numbers" result $\overline{ \Delta N^2} \propto\overline{N}$; instead, we find
\begin{equation}
\overline{ \Delta N^2 } \propto \bar{N}^{2\phi(d)} \ ,
\label{gnf}
\end{equation}
with the exponent $\phi(d)$ given not by the ``law of large numbers" result
$\phi=1/2$, but, rather,
\begin{eqnarray}
\phi(d)&=&\left\{
\begin{array}{ll}
1/2+1/d\,\,\,\,\,\,\,\,\,,
&\gvt<0\,\,\rm{or} \,\, d=2 \,,
\\ \\
1+\chi/d\,\,\,\,\,\,\,\,\,,
&\gvt>0\,\,\rm{and} \,\, d>2 \ .
\end{array}\right.
\label{phi}
\end{eqnarray}


The remainder of this paper is organized as follows.  In the next section, we derive a hydrodynamic model for flocks with quenched noise. We study the hydrodynamic model to linear order in fluctuations about a state of perfect order in section III. In section IV, we present the full nonlinear theory for the four cases: A) $\gamma v_2 <0$, $d>2$; B) $\gamma v_2 <0$, $d=2$; C) $\gamma v_2 >0$, $d=2$; D) $\gamma v_2 >0$, $d>2$. In section V,  we describe a numerical model to study flocking with quenched disorder. The results from our numerical studies are presented in section VI, before we conclude in section VII.

\section{The Hydrodynamic Model}\label{The Hydrodynamic Model}



Our starting point is the hydrodynamic theory of \cite{TT1,TT2,TT3,TT4,NL}, modified only by the inclusion of a quenched random force {\bf f}. In the ordered phase, this takes the form of the following pair of coupled equations of motion for the fluctuation ${\bf v}_{\perp}(\br,t)$ of the local velocity of the flock perpendicular to the direction of mean flock motion (which mean direction will hereafter denoted as "$\parallel$"), and the departure $\delta\rho(\br,t)$ of the density from its mean value $\rho_0$:

\begin{widetext}
\begin{eqnarray}
&\partial_{t} {\bf v}_{\perp} + \gamma\partial_{\parallel} 
{\bf v}_{\perp} + \lambda \left({\bf v}_{\perp} \cdot
{\bf \nabla}_{\perp}\right) {\bf v}_{\perp} =-g_1\delta\rho\partial_{\parallel} 
{\bf v}_{\perp}-g_2{\bf v}_{\perp}\partial_{\parallel}
\delta\rho-g_3{\bf v}_{\perp}\partial_t
\delta\rho -{c_0^2\over\rho_0}{\bf \nabla}_{\perp}
\delta\rho -g_4{\bf \nabla}_{\perp}(\delta \rho^2)\nonumber\\&+
D_B{\bf \nabla}_\perp\left({\bf \nabla}_\perp\cdot{\bf v}_\perp\right)+
D_T\nabla^{2}_{\perp}{\bf v}_{\perp} +
D_{\parallel}\partial^{2}_{\parallel}{\bf v}_{\perp}+\nu_t\partial_t{\bf \nabla}_{\perp}\delta\rho+\nu_\parallel\partial_\parallel{\bf \nabla}_{\perp}\delta\rho+{\bf f}_{\perp} ~, 
\label{vEOMbroken}\\
&\partial_t\delta
\rho +\rho_o{\bf \nabla}_\perp\cdot{\bf v}_\perp
+\lambda_{\rho}{\bf \nabla}_\perp\cdot({\bf v}_\perp\delta\rho)+v_2
\partial_{\parallel}\delta
\rho =D_{\rho\parallel}\partial^2_\parallel\delta\rho+D_{\rho v} \partial_{\parallel}
\left({\bf \nabla}_\perp \cdot {\bf v}_{\perp}\right)+\phi\partial_t\partial_\parallel\delta\rho
+w_1 \partial_\parallel(\delta\rho^2)+w_2\partial_\parallel(|
{\bf v}_\perp|^2)~, \nonumber \\
\label{cons broken}
\end{eqnarray}
\end{widetext}
where  $\gamma$, $\lambda$,  $\lambda_\rho$,
$c_0^2$, $g_{1,2,3,4}$, $w_{1,2}$,  $D_{B\rm{eff},T,\parallel, \rho\parallel, \rho v}$, $\nu_{t,\parallel}$, $v_2$, $\phi$,  and $\rho_0$  are all phenomenological constants, the last being  the mean density.

In flocks {\it without} quenched disorder \cite{TT1,TT2,TT3,TT4,NL}, the random force  ${\bf f}({\bf r}, t)$ is taken to be a Langevin noise, uncorrelated in  {\it both} space  {\it and} time. To treat quenched disorder, we simply take this random force to be {\it static}; i.e., to  depend {\it only} on position: ${\bf f}({\bf r},t)={\bf f}({\bf r})$, and not on time $t$ at all, with short-ranged spatial  correlations:

\begin{equation}
 \overline{ f^\perp_i ({\bf r}) f^\perp_j ({\bf r}^{~\prime})} = \Delta \delta^\perp_{ij} \delta^d ({\bf r}-{\bf r}^{~\prime})
\label{fcor}
\end{equation}
where we use an overbar to denote averages over the quenched disorder, and $\delta^\perp_{ij}=1$ if and only if $i=j\ne\parallel$, and is zero for all other $i$, $j$.  We will also assume ${\bf f_\perp}$ is zero mean, and Gaussian. Including a Langevin (i.e., time-dependent) component in addition to this quenched force (which we actually do in our simulations) changes none of the  results presented here, since it is subdominant relative to the quenched disorder (although it can change {\it time-dependent} correlations, as we will discuss in a future publication \cite{future}). Small departures from Gaussian statistics can be shown to be irrelevant in the renormalization group sense.

\section{Linear Theory}\label{Linear Theory}

We first analyze the hydrodynamic model by linearizing the equations in $\delta\rho$ and $\vp$, which gives: 
\begin{widetext}
\begin{eqnarray}
&\partial_{t} {\bf v}_{_\perp} + \gamma\partial_{{_\parallel}}
{\bf v}_{_\perp} = -{c_0^2\over\rho_0}{\bf \nabla}_{_\perp}
\delta\rho+
D_B{\bf \nabla}_{_\perp}\left({\bf \nabla}_{_\perp}\cdot{\bf v}_{_\perp}\right)+
D_T\nabla^{2}_{_\perp}{\bf v}_{_\perp} +
D_{{_\parallel}}\partial^{2}_{{_\parallel}}{\bf v}_{_\perp}+\nu_t\partial_t{\bf \nabla}_{_\perp}\delta\rho+\nu_{_\parallel}\partial_{_\parallel}{\bf \nabla}_{_\perp}\delta\rho+{\bf f}_{\perp}\\
\label{vEOMbrokenlin}
&\partial_t\delta
\rho +\rho_o{\bf \nabla}_{_\perp}\cdot{\bf v}_{_\perp}
+v_2
\partial_{{_\parallel}}\delta
\rho =D_{\rho{_\parallel}}\partial^2_{_\parallel}\delta\rho+D_{\rho{_\perp}}\nabla^2_{_\perp}\delta\rho+D_{\rho v} \partial_{{_\parallel}}
\left({\bf \nabla}_{_\perp} \cdot {\bf v}_{_\perp}\right)+\phi\partial_t\partial_{_\parallel}\delta\rho
\label{cons brokenlin}
\end{eqnarray}
\end{widetext}
The steady-state solution of these equations is readily obtained by taking ${\bf \delta v}({\bf r},t)$ and $\delta\rho({\bf r},t)$ to be time independent, and spatially Fourier transforming the resultant equations. This gives a set of linear equations relating ${\bf \delta v}({\bf q})$ and $\delta\rho({\bf q})$ to the corresponding spatial Fourier transforms of  the quenched random force ${\bf f}({\bf q})$:

\begin{equation}
\left[i \rho_0+D_{\rho v} q_\parallel\right] q_\perp v_L +\left[i v_2 q_\parallel +\Gamma_{\rho}({\bf q})\right] \delta\rho  =0,
\label{steadystaterho}
\end{equation}
\begin{equation}
\left[i \gamma q_\parallel +\Gamma_L({\bf q})\right] v_L +\left[ {ic_0^2\over\rho_0}-\nu_\parallel q_\parallel \right]q_\perp\delta\rho  =f_L({\bf q}),
\label{steadystateVL}
\end{equation}
\begin{equation}
\left[i \gamma q_\parallel +\Gamma_T({\bf q})\right] {\bf v}_T   ={\bf f}_T({\bf q}),
\label{steadystateVT}
\end{equation}
where we've defined the wavevector-dependent dampings

\begin{eqnarray}
&\Gamma_\rho({\bf q})  \equiv D_{\rho\parallel} q_\parallel^2+D_{\rho\perp}q_\perp^2~,
\label{Gammarhodef}\\
&\Gamma_L({\bf q})  \equiv D_{\parallel} q_\parallel^2+D_{\perp}q_\perp^2,
\label{GammaLdef}\\
&\Gamma_T({\bf q})  \equiv D_{\parallel} q_\parallel^2+D_Tq_\perp^2~,
\label{GammaTdef}
\end{eqnarray}
all of which vanish like $q^2$ as $q\rightarrow 0$. Here we've defined $D_\perp\equiv D_B+D_T$,
and we've also separated the velocity ${\bf v}_{_\perp}$ and the noise ${\bf f}_\perp$ into components $f_L({\bf q})$ along and ${\bf f}_T({\bf q})$ perpendicular to the projection ${\bf q}_{_\perp}$
of ${\bf q}$  perpendicular to $<{\bf v}>$ via
\begin{eqnarray}
v_L\equiv {\bf v}_{_\perp}\cdot{\bf q}_{_\perp}/q_{_\perp}
~,~{\bf v}_T\equiv {\bf v}_{_\perp}- v_L {{\bf q}_{_\perp}\over q_{_\perp}}~,
\label{vLTdef}
\end{eqnarray}
with $f_L$ and ${\bf f}_T$ obtained from ${\bf f}$ in the same way. Note that ${\bf v}_T$ is identically zero in $d=2$, since ${\bf q}_{_\perp}$ has only one non-zero component in that dimension.

Equations (\ref{steadystateVL}-\ref{steadystateVT}) are a simple set of linear algebraic equations for the velocity and density fluctuations $\delta\rho$, $v_L$, and ${\bf v}_T$,  which can easily be solved for these fields in terms of the noises $f_L$ and ${\bf f}_T$. We find:
\begin{eqnarray}
&\delta\rho({\bf q})  =G_\rho({\bf q})f_L({\bf q})~,
\label{rhosol}\\ 
&v_L({\bf q})  =G_L({\bf q})f_L({\bf q})~,
\label{vLsol}\\ 
&{\bf v}_T({\bf q})  =G_T({\bf q}){\bf f}_T({\bf q})~,
\label{vTsol}
\end{eqnarray}
where the ``propagators" $G_\rho, {L, T}({\bf q})$ are given, dropping ``irrelevant" terms (i.e., terms that are higher order in $q$), by:
\begin{equation}
G_\rho({\bf q})={-i\rho_0q_\perp\over c_0^2q_\perp^2-\gamma v_2q_\parallel^2+i\Xi({\bf q})q_\parallel}~,
\label{Grhodef}
\end{equation}
\begin{equation} 
G_L({\bf q})={iv_2q_\parallel\over c_0^2q_\perp^2-\gamma v_2q_\parallel^2+i\Xi({\bf q})q_\parallel}~,
\label{GLdef}
\end{equation}
\begin{equation}
G_T({\bf q})={1\over \left[i \gamma q_\parallel +\Gamma_T({\bf q})\right] }~,
\label{GTdef}
\end{equation}
where we've defined another wavevector dependent damping
\begin{equation}
\Xi({\bf q})=\gamma\Gamma_\rho+v_2\Gamma_L+\left(\nu_\parallel\rho_0-{c_0^2D_{\rho v}\over\rho_0}\right)q_\perp^2
=Aq_\perp^2+Bq_\parallel^2~,
\label{Xidef}
\end{equation}
which also scales like $q^2$ as $q\rightarrow 0$. In Eq.~(\ref{Xidef}), we've defined
\begin{eqnarray}
A&\equiv&
\left[v_2D_\perp+\gamma D_{\rho\perp}+\nu_\parallel\rho_0-{c_0^2D_{\rho v}\over\rho_0}\right]~,
\nonumber\\
B&\equiv&
\left[v_2D_\parallel+\gamma D_{\rho\parallel}\right]~.
\label{ABdef}
\end{eqnarray}
We can now obtain the disorder averaged correlation functions of the fluctuations $ \overline{|v_L({\bf q})|^2}$, $ \overline{|\rho({\bf q})|^2}$, and $ \overline{|{\bf v}_T({\bf q})|^2}$ simply by using Eqs.~(\ref{rhosol}--\ref{vTsol}) to relate these averages to the noise averages Eq.~(\ref{fcor}).
This gives:
\begin{equation}
\overline{|v_L({\bf q})|^2}={v_2^2q_\parallel^2\Delta\over \left[c_0^2q_\perp^2-\gamma v_2q_\parallel^2\right]^2+\Xi({\bf q})^2q_\parallel^2}~,
\label{vLqfluc}
\end{equation}
\begin{equation}
\overline{|\delta\rho({\bf q})|^2}={\rho_0^2q_\perp^2\Delta\over \left[c_0^2q_\perp^2-\gamma v_2q_\parallel^2\right]^2+\Xi({\bf q})^2q_\parallel^2}~,
\label{rhoqfluc}
\end{equation}
and
\begin{equation}
\overline{|{\bf v}_T({\bf q})|^2}={(d-2)\Delta\over \gamma^2 q_\parallel^2 +\Gamma_T({\bf q})^2}~,
\label{vTqfluc}
\end{equation}
where $d$ is the dimension of space.

These expressions can be rewritten 
in terms of the magnitude $q$ of wavevector $\bq$ and the angle $\theta_q$ between \xpa and $\bq$:
\begin{equation}
\overline{|v_L({\bf q})|^2}= \frac{\tilde{\Delta}\cos^2\theta_{{\bf q}}}{q^2[\epsilon^2(\theta_{{\bf q}})q^2 + (\sin^2\theta_{{\bf q}}-\left[{\gamma v_2\over c_0^2}\right] \cos^2\theta_{{\bf q}})^2]},
\label{vLanglefluc}
\end{equation}
\begin{equation}
\overline{|\delta\rho({\bf q})|^2}=\frac{\tilde{\Delta}(\rho_0^2/v_2^2)\sin^2\theta_{{\bf q}}}{q^2[\epsilon^2(\theta_{{\bf q}})q^2 + (\sin^2\theta_{{\bf q}}-\left[{\gamma v_2\over c_0^2}\right]\cos^2\theta_{{\bf q}})^2]}~,
\label{rhoanglefluc}
\end{equation}
and
\begin{equation}
\overline{|{\bf v}_T({\bf q})|^2}={(d-2)\Delta\over \gamma^2q^2\left[\epsilon_T^2(\theta_{{\bf q}})q^2 +  \cos^2\theta_{{\bf q}}\right]}~,
\label{vTanglefluc}
\end{equation}
where we've defined  $\tilde\Delta\equiv{v_2^2\Delta\over c_0^4}$,
and direction-dependent damping coefficients
$\epsilon(\theta_{{\bf q}})\equiv\Xi({\bf q})/ c_0^2q^2=(A\cos^2\theta_{{\bf q}}+B\sin^2\theta_{{\bf q}} )/c_0^2$ and $\epsilon_T(\theta_{{\bf q}})\equiv\Gamma_T({\bf q})/ \gamma^2q^2=(D_\parallel\cos^2\theta_{{\bf q}}+D_T\sin^2\theta_{{\bf q}} )/\gamma^2$. 

From Eqs.~(\ref{vLanglefluc}--\ref{vTanglefluc}), we immediately see that there is an important distinction between the cases  $\gvt>0$ and $\gvt<0$. In the former case,  fluctuations of $v_L$ and $\rho$ are highly anisotropic: 
they scale like $q^{-2}$ for all directions of ${\bf q}$ {\it except}  when $\theta_{{\bf q}}=\theta_c$ or $\pi-\theta_c$,  where we have defined a critical angle of propagation $\theta_c\equiv\arctan \left[{\sqrt{\gamma v_2}\over c_0}\right]$.  The physical significance of $\theta_c$ is that it is the direction in which the speed of propagation of longitudinal sound waves in the flock vanishes \cite{TT1,TT2,TT3,TT4,NL}.
 For these special directions (which only exist if $\gvt>0$) both $\overline{|v_L({\bf q})|^2}$ and $\overline{|\delta
\rho({\bf q})|^2}$ scale like $q^{-4}$. On the other hand, when $\gvt<0$, fluctuations of $v_L$ and $\rho$ are essentially isotropic: 
they scale as $q^{-2}$ for {\it all} directions of ${\bf q}$ full stop.


Fluctuations of $\vt$, however, are {\it always} anisotropic, diverging as $q^{-4}$ for $\theta_{{\bf q}}=\pi/2$, and as $q^{-2}$ for all other directions of $\bq$. Of course, there {\it are} no such fluctuations in $d=2$, since, as noted earlier, $\vt$ does not exist in that case, as reflected by the factor of $(d-2)$ in Eq.~(\ref{vTanglefluc}).


It is intuitively clear why the fluctuations are so much larger for ${\bf q}$ in these special directions that exist in the case $\gamma v_2>0$. The longitudinal degrees of freedom $v_L$ and $\delta\rho$ are carried by propagating sound waves in the flock \cite{TT1,TT2,TT3,TT4, NL}. For directions in which these sound waves have a non-vanishing speed, the static disorder looks, in a frame co-moving with the sound mode, like a time dependent one, which quickly time averages to zero. Fluctuations in these directions are therefore small, and are regulated by the sound speeds, which involve $c_0$ and $v_2$. For the special directions $\theta_{{\bf q}}=\theta_c$, however, the sound speeds vanish \cite{TT1,TT2,TT3,TT4,NL}, and so these fluctuations sit right on top of the static quenched disorder, and grow until cutoff by the damping, which is higher order in $q$ than sound propagation. The entire phenomenon is similar to a very under-damped oscillator: when driven off resonance, the response is small, and almost independent of the damping, while on resonance,
 the response is large, and controlled entirely by the damping. Here the resonance is at zero frequency, which is achieved by varying the direction of propagation, but the underlying physics is exactly the same.

The same argument applies for the transverse fluctuations Eq.~(\ref{vTanglefluc}), only now the critical angle at which the propagation speed of these modes vanishes is \cite{TT1,TT2,TT3,TT4,NL} $\theta_{c, \rm{transverse}}=\pi/2$.

The above discussion
assumed that
$\gamma$ and $v_2$ have the same sign. While this has always proved to be the case in the few systems for which $\gamma$ and $v_2$ have been deduced from simulations \cite{TT2} (including the  simulations we report here), there is no symmetry argument that this must always be true. We therefore expect there to be some flocking systems in which these parameters have opposite signs. In this case,  there is no (real) $\theta_c$, and both $\overline{|v_L({\bf q})|^2}$ and $\overline{|\delta
\rho({\bf q})|^2}$ scale like $1/q^2$ for {\it all} directions of ${\bf q}$. As a result, {\it only} the fluctuations of ${\bf v}_T({\bf q}) $ become anomalously large for some directions of propagation; namely $\theta=\theta_{c, \rm{transverse}}=\pi/2$, as before.

Of course, in $d=2$, as noted earlier, there {\it is} no transverse component of ${\bf v}$. Hence, in $d=2$, when $\gamma v_2<0$, both the {\it total} mean squared velocity fluctuation $\overline{|{\bf v}({\bf q})|^2}$ and the mean squared density fluctuation
$\overline{|\delta\rho({\bf q})|^2}$ scale like $1/q^2$ for {\it all} directions of ${\bf q}$. While we've derived this result in the linearized theory, it proves to hold in the full theory as well.

In higher spatial dimensions $d>2$, the transverse fluctuations now exist, and are still soft ( with a critical angle $\theta_c=\pi/2$), even when $\gamma$ and $v_2$ have opposite signs. We will see later, however, that they are {\it not}   as soft as the linear theory predicts.

In any event, the distinction between the cases  $\gamma v_2>0$ and $\gamma v_2<0$ is significant, even in the linearized theory. It becomes even more relevant in the non-linear theory; as we will show below, it is possible to make a compelling argument giving exact exponents for all spatial dimensions $d\le 5$ in the case $\gamma v_2<0$, but not for $\gvt>0$.

Returning now to the case $\gamma v_2>0$, we see that it is
the special directions $\theta_{{\bf q}}$ near $\theta_c$ and ${\pi\over 2}$ of wavevector ${\bf q}$ that dominate the real space fluctuations $\overline{|{\bf v}_\perp({\bf r})|^2}$  and $\overline{|\delta\rho({\bf r})|^2}$. These  can be obtained by integrating the Fourier transformed fluctuations
$\overline{|\delta\rho({\bf q})|^2}$, $\overline{|v_L({\bf q})|^2}$, and $\overline{|{\bf v}_T({\bf q})|^2}$ over all wavevector ${\bf q}$. Focusing for now on the real space {\it velocity} fluctuations, which determine whether or not the system exhibits long ranged orientational order (i.e., a non-zero average velocity $\overline{{\bf v}({\bf r})}$), and expanding
$\overline{|v_L({\bf q})|^2}$  and $\overline{|{\bf v}_T({\bf q})|^2}$ for $\theta_{{\bf q}}$ near $\theta_c$ and ${\pi\over 2}$
 respectively,  we obtain
\begin{equation}
\overline{|v_L({\bf q})|^2}\approx{\Delta v_2^2 g_L(q, \delta\theta)\over c^4_0 q^2}
\label{vLexp}
\end{equation}
where we've defined $\delta\theta\equiv\theta_{{\bf q}}-\theta_c$ and
\begin{equation}
g_L(q, \delta\theta)={\cos^2\theta_c\over m \delta\theta^2+\aleph^2 q^2}~,
\label{gLdef}
\end{equation}
with the constants $m$ and $\aleph$ given by $m={4v_2\gamma\over c_0^2}$ and

\begin{eqnarray}
c_0^2 \aleph &&=\left(v_2D_\perp+\gamma D_{\rho\perp}+\nu_\parallel\rho_0-{c_0^2D_{\rho v}\over\rho_0}\right)\sin^2\theta_c\cos\theta_c \nonumber\\
&&+\left(\gamma D_{\rho\parallel}+v_2 D_\parallel\right)\cos^3\theta_c~,
\label{alephdef}
\end{eqnarray}

and

\begin{equation}
\overline{|{\bf v}_T({\bf q})|^2}\approx{(d-2)\Delta  g_T(q, \delta\theta_2)\over \gamma^2 q^2}
\label{vTexp}
\end{equation}
where we've defined $\delta\theta_2\equiv\theta_{{\bf q}}-{\pi\over 2}$ and
\begin{equation}
g_T(q, \delta\theta_2)={1\over  \delta\theta_2^2+\beta^2 q^2}~,
\label{gTdef}
\end{equation}
with the constant $\beta$ given by $\beta={D_T\over \gamma}$.

Using these approximations to evaluate the real space fluctuations $\overline{|{\bf v}_\perp({\bf r})|^2}$ gives

\begin{gather}
\begin{gathered}
\overline{|{\bf v}_\perp({\bf r})|^2}=\int{d^dq\over(2\pi)^d}
\left(\overline{|v_L({\bf q})|^2}
+ \overline{|{\bf v}_T({\bf q})|^2}\right) \\
\approx\int{q^{d-3}dq\over(2\pi)^d}
\left({\Delta v_2^2 \over c^4_0}
\left[\int_{-\infty}^{\infty}{\sin^{d-2}\theta_cd\delta\theta}~g_L(q, \delta\theta)\right] \right. \\
\left.+{(d-2)\Delta \over \gamma^2_0 q^2}\left[\int_{-\infty}^{\infty}{d\delta\theta_2}g_T(q, \delta\theta_2)\right]\right)
\label{vrfluc}
\end{gathered}
\end{gather}
where we've used the fact (which will become evident in a moment) that, for small $q$, the angular integrals are dominated by $\delta\theta<<1$ and $\delta\theta_2<<1$  to extend the range of those integrals to $\pm\infty$.

Evaluating those angular integrals is straightforward, and gives:
\begin{gather}
\begin{gathered}
\int_{-\infty}^{\infty}{\sin^{d-2}\theta_cd\delta\theta}~g_L(q, \delta\theta)=\\
\int_{-\infty}^{\infty}{\sin^{d-2}\theta_cd\delta\theta}~{\cos^2\theta_c\over m^2 \delta\theta^2+\aleph^2 q^2}=\\
{\pi\sin^{d-2}\theta_c\cos^2\theta_c\over m\aleph q}\propto 1/q
\label{angintL}
\end{gathered}
\end{gather}

\begin{gather}
\begin{gathered}
\int_{-\infty}^{\infty}{d\delta\theta_2}g_T(q, \delta\theta_2)=\\
\int_{-\infty}^{\infty}{d\delta\theta_2}~{1\over  \delta\theta_2^2+\beta^2 q^2}=
{\pi\over \beta q}\propto 1/q
\label{angintT}
\end{gathered}
\end{gather}

Inserting these results back into Eq.~(\ref{vrfluc}), we see that the linearized theory predicts that
\begin{equation}
\overline{|{\bf v}_\perp({\bf r})|^2}\propto\int q^{d-4}dq~,
\label{vrflucscale}
\end{equation}
which clearly diverges in the long wavelength (i.e., infra-red, or $q\rightarrow 0$) limit for $d\le d^{lin}_c=3$. This implies that, {\it according to the linearized theory}, there should be no long-ranged orientational order for $d\le d^{lin}_c=3$; that is, the ordered flock, with a nonzero $\overline{{\bf v}({\bf r})}$, should not occur for $d\le 3$, {\it no matter how weak the disorder}. In the critical dimension $d=3$, quasi-long-ranged order (with algebraic decay of velocity correlations in space), should, again according to the {\it linearized} theory, occur.

For the case $\gamma v_2<0$, in $d=2$, the fluctuations predicted by the linear theory are far smaller, due to the absence of $\vt$, and the fact that for the only remaining velocity fluctuations- namely, the longitudinal ones $v_L$- no longer 
there are {\it no} directions of $\bq$ in which the linear theory predicts a divergence of $\vq$ stronger than $1/q^2$ as $q\to0$. As a result, for $\gamma v_2<0$, in $d=2$, we have 
\begin{equation}
\overline{|{\bf v}_\perp({\bf r})|^2}\propto\int {dq\over q}~,
\label{vrflucscale2}
\end{equation}
which implies only a logarithmic divergence of velocity fluctuations, in contrast to the strong (algebraic) divergence found above for the $\gamma v_2>0$ case.

We will see in the next section that, when non-linearities are taken into account, these two cases become much more similar, with {\it both}  exhibiting 
 quasi-long-ranged order Eq.~(\ref{qlro}).

\section{Nonlinear Theory}\label{Nonlinear Theory}
The large fluctuations in this system lead one to worry about the validity of the linear approximation just presented. This worry is, in fact, justified: we now show that the non-linearities explicitly displayed in Eqs.~(\ref{vEOMbroken},\ref{cons broken}) in fact radically change the scaling of fluctuations in flocks with quenched disorder for all spatial dimensions $d\le5$. Furthermore, this change in scaling in fact stabilizes long-ranged orientational order (i.e., makes it possible for the flock to acquire a non-zero mean velocity
($<{\bf v}>\ne 0$)) in three dimensions, and may make quasi-long-ranged order possible in two dimensions.
Here, as usual, by  ``quasi-long-ranged order" we  mean real space velocity correlations that decay algebraically with separation; i.e., Eq.~(\ref{qlro}).

As we saw in our linearized analysis, our problem qualitatively changes when $\gamma v_2$ changes sign. For $\gamma v_2>0$, there are directions of ${\bf q}$ for which the longitudinal sound speeds vanish, and, hence, the longitudinal modes make appreciable contributions to the fluctuations. In contrast, for  $\gamma v_2<0$, there are  no such directions of ${\bf q}$, and so the contributions of longitudinal modes can be neglected relative to those of the transverse modes, {\it except}, of course, in $d=2$, where there {\it are} no tranverse modes. 
 Hence, there are four distinct cases we must analyze separately: A) $\gamma v_2<0$, $d>2$; B) $\gamma v_2<0$, $d=2$; C) $\gamma v_2>0$, $d=2$; D) $\gamma v_2>0$, $d>2$. We will now discuss the behavior of the full, non-linear theory for each of these four cases in turn.


\subsection{$\gamma v_2<0$, $d>2$}

\subsubsection{Linear scaling, and relevance of non-linearities for $d\le 5$}

We begin by demonstrating that the aforementioned non-linearities become important for spatial dimensions $d\le 5$. We do this first for the case $\gamma v_2<0$, for which,
we remind the reader, the sound speeds do not vanish for any direction of propagation.

We can assess the importance of the non-linearities by  power counting on the 
equations of motion Eqs.~(\ref{vEOMbroken},\ref{cons broken}). This power counting is quite subtle, due to the  anisotropy of the fluctuations in this system. We will accordingly rescale coordinates $r_\parallel$ along the  direction of flock motion differently from
those ${\bf r_\perp}$ orthogonal to that direction, taking the rescaling factor for ${\bf r_\perp}$ to be $b$, and that for $r_\parallel$ to be $b^\zeta$, where  the anisotropy exponent $\zeta$ is  to be determined. 

To complete the rescaling, we will also rescale time by a factor of $b^z$, where $z$ is known as the ``dynamical exponent", and the fields ${\bf v}_\perp$ and $\rho$ by  factors of $b^\chi$ and $b^{\chi_\rho}$; where $\chi$ and $\chi_\rho$ are the ``roughness exponents" for ${\bf v}_\perp$ and $\rho$ respectively. We choose the same rescaling factor for the fields ${\bf v}_\perp$ and $\rho$ because, as we saw in our treatment of the linearized model, their fluctuations scale in the same way with wavevector.

To summarize, our rescaling is as follows:
\begin{eqnarray}
{\bf r}_\perp &\rightarrow& b {\bf r}_\perp ~~, \nonumber \\
r_\parallel&\rightarrow& b^\zeta r_\parallel ~~, \nonumber \\
t &\rightarrow& b^z t ~~, \nonumber \\
{\bf v}_\perp &\rightarrow& b^\chi {\bf v}_\perp ~~, \nonumber \\
\delta \rho &\rightarrow& b^{\chi_\rho} \delta \rho ~~.
\label{rescale}
\end{eqnarray}

Performing these rescalings as just described, we easily find how  the parameters in the rescaled equations (denoted by primes) are related to those of the unrescaled equations. We will focus on those parameters that actually affect the fluctuations in the dominant regime of wavevector ${\bf q}$, which, as we noted in the linearized section, is the regime $q_\parallel\sim q_\perp^2\ll q_\perp$ (this being the regime in which $\overline{{\bf v}_T^2}\propto1/q^4\gg\overline{v_L^2}\propto1/q^2$). Inspection of our expressions Eqs.~(\ref{rhoqfluc}, \ref{vLqfluc}, \ref{vTqfluc}) for the fluctuations of the density $\rho$ and the longitudinal and transverse velocities $v_{L,T}$ shows that in this regime of wavevector, the fluctuations are entirely determined (in the linearized approximation) by the parameters $\Delta$, $\gamma$, and $D_T$, and the combination of parameters ${c_0^2\over\rho_0}$. (Note that in the wavevector limit we are considering, both the $\Xi({\bf q})^2 q_\parallel^2$ and the $\gamma v_2 q_\parallel^2$ terms in the denominators of Eqs.~(\ref{rhoqfluc}, \ref{vLqfluc}), as well as the $D_\parallel$ term in the denominator of Eq.~(\ref{vTqfluc})
are negligible, and can be dropped with impunity). We will therefore focus on the rescaling of these parameters under Eq.~(\ref{rescale}), which are easily found to be given by:

\begin{eqnarray}
\gamma^{\prime}&=&b^{z-\zeta}\gamma  ~~,
\label{gammarescale}
\end{eqnarray}
\begin{eqnarray}
\left({c_0^2\over\rho_0}\right)^{\prime}&=&b^{{\chi_\rho-\chi+z-1}}\left({c_0^2\over\rho_0}\right)  ~~,
\label{c_0rescale}
\end{eqnarray}
\begin{eqnarray}
D_T^{\prime}&=&b^{z-2}D_T  ~~,
\label{Drescale}
\end{eqnarray}
\begin{eqnarray}
\Delta^{\prime}&=&b^{2(z-\chi)+1-d-\zeta}\Delta ~~.
\label{Deltarescale}
\end{eqnarray}
We can thus keep the scale of the fluctuations of $\rho$ and $v_L$ fixed by choosing the exponents $z$, $\zeta$, $\chi$, and $\chi_\rho$ to obey
\begin{eqnarray}
z-\zeta=0  ~~,
\label{gammafix}
\end{eqnarray}
\begin{eqnarray}
\chi_\rho-\chi+z-1=0  ~~,
\label{c_0fix}
\end{eqnarray}
\begin{eqnarray}
z-2=0 ~~,
\label{Dfix}
\end{eqnarray}
\begin{eqnarray}
2(z-\chi)+1-d-\zeta=0 ~~.
\label{Deltafix}
\end{eqnarray}
This system of linear equations can readily be solved for all of the exponents, yielding
\begin{eqnarray}
z_{\rm{lin}}=\zeta_{\rm{lin}}=2 ,~\chi_{\rm{lin}}={3-d\over2},~\chi_{\rho,\rm{lin}}={1-d\over2}.
\label{linexp}
\end{eqnarray}
The subscript ``lin" in these expressions denotes the fact that we have determined these exponents ignoring the effects of the non-linearities in the equations of motion Eqs.~(\ref{vEOMbroken},\ref{cons broken}). We now use them to determine in what spatial dimension $d$ those non-linearities become important.

Upon the rescalings Eq.~(\ref{rescale}), the  non-linear terms
$\lambda$,  and
$g_{1,2,3,4}$ in the $\vp$ equation of motion Eq.~(\ref{vEOMbroken})  obey
\begin{eqnarray}
\lambda^{\prime}&=&b^{z+\chi-1}\lambda= b^{5-d\over2}\lambda ~~,
\label{lambdarescale}
\end{eqnarray}

\begin{eqnarray}
g_1^{\prime}&=&b^{z+\chi_\rho-\zeta}g_1 =b^{1-d\over2}g_1 ~~,
\label{g1rescale}
\end{eqnarray}

\begin{eqnarray}
g_2^{\prime}&=&b^{z+\chi_\rho-\zeta}g_2 =b^{1-d\over2}g_2 ~~,
\label{g2rescale}
\end{eqnarray}

\begin{eqnarray}
g_3^{\prime}&=&b^{\chi_\rho}g_3 =b^{1-d\over2}g_3 ~~.
\label{g3rescale}
\end{eqnarray}

\begin{eqnarray}
g_4^{\prime}&=&b^{z+2\chi_\rho-\chi-1}g_4 =b^{1-d\over2}g_4 ~~.
\label{g4rescale}
\end{eqnarray}

The behavior of these rescaled parameters for large rescaling factor $b$ tells us which parameters are important at long distances (namely, those that grow with increasing $b$). We'd like to assess the importance of the nonlinear terms. Since we have chosen our rescaling factors Eq.~(\ref{linexp}) to keep the size of the fluctuations  $v_{L,T}$ and $\delta\rho$ fixed, we can directly assess whether or not the nonlinear terms grow in importance by whether or not their coefficients $\lambda$ do (since the factors involving the fields in those terms will not change upon rescaling). 

By inspection of Eqs.~(\ref{lambdarescale}--\ref{g4rescale}),
we see that only $\lambda$ becomes relevant in any spatial dimension $d>1$; in fact, it becomes relevant for $d\le d_c=5$. We will now discuss the implications of this in the next subsection.
\\

\subsubsection{Non-linear scaling for $d \le 5$}

The results so far imply that for all spatial dimensions $d\le 5$, we must keep the $\lambda$ non-linearity in Eq.~(\ref{vEOMbroken}), but can drop the $g_{1,2,3,4}$ non-linearities.  Furthermore, if we restrict ourselves to consideration of the transverse modes $\vt$, which we can do by projecting 
the spatial Fourier transform of Eq.~(\ref{vEOMbroken}) perpendicular to $\qp$, we see that there is {\it no} coupling between $\vt$ and $\rho$ {\it at all}, even at nonlinear order. Hence, $\rho$ completely drops out of the problem of determining the fluctuations of $\vt$. And since $\vt$ is, as we saw in our treatment of the linearized version of this problem, the dominant contribution to the velocity fluctuations when $d>2$ (so that $\vt$ actually exists) and $\gamma v_2<0$ (so that there is no direction of $\bq$ for which the {\it longitudinal} velocity fluctuations $v_L$ diverge more strongly than $1/q^2$ in the linearized approximation), this means that the long distance scaling of the velocity fluctuations will be the same as in a model with no density fluctuations {\it at all}; that is, an incompressible model. 

Such a model takes the form
\begin{widetext}
\begin{eqnarray}
\partial_{t} {\bf v}_{\perp} + \gamma\partial_{\parallel} 
{\bf v}_{\perp} + \lambda \left({\bf v}_{\perp} \cdot
{\bf \nabla}_{\perp}\right) {\bf v}_{\perp} =-{\bf \nabla}_{\perp}
P +
D_T\nabla^{2}_{\perp}{\bf v}_{\perp} +
D_{\parallel}\partial^{2}_{\parallel}{\bf v}_{\perp}+{\bf f}_{\perp}  \ ,
\label{incompEOM}
\end{eqnarray}
\end{widetext}
with the pressure $P$ determined {\it not} by the density $\rho$, but by the incompressibility condition 
\beq
\nabla_\perp\cdot\vp=0 \ .
\label{incompcond}
\eeq

This equation of motion has a number of useful properties that make it possible for us to determine the scaling laws exactly. These are:

\noindent 1) The only nonlinearity (the $\lambda$ term) can  be written as a total $\perp$-derivative. This follows from the vector calculus identity:
\begin{eqnarray}
\left({\bf v}_{_\perp} \cdot
{\bf \nabla}_{_\perp}\right) v^{\perp}_i=\partial^\perp_j\left(v^\perp_jv^\perp_i\right)-v^\perp_i{\bf \nabla}_{_\perp}\cdot{\bf v_{_\perp}}\,\,.
\label{trans1}
\end{eqnarray}

The first term on the right hand side of this expression is obviously a total $\perp$-derivative. The second term vanishes by the incompressibility condition Eq.~(\ref{incompcond}).

This implies that the nonlinearity can {\it only}
renormalize terms which themselves involve $\perp$-derivatives (i.e., $D_T^0$); specifically, there are {\it no} graphical corrections to either $\gamma$ or $\Delta$.

\noindent 2)  There are no graphical corrections
$\lambda$ either, because the equation of motion  Eq.~(\ref{incompEOM}) has  an exact
``pseudo-Galilean invariance" symmetry: that is, it remains
unchanged by
a pseudo-Galilean
the transformation:
\begin{eqnarray}
{\bf r}_\perp \to {\bf r}_\perp-\lambda {\bf v}_1 t~~~,~~
{\bf v}_{_\perp} \to {\bf v}_{_\perp} + {\bf v}_1~~~
,\label{Gal}
\end{eqnarray}
 for arbitrary constant vector ${\bf v}_1\perp\hat{x}_{_\parallel}$.
Note that if $\lambda=1$, this reduces to the familiar Galilean invariance in the $\perp$-directions.
Since such an exact symmetry  must continue to hold upon renormalization, with the {\it same} value of  $\lambda$,  the parameter $\lambda$ cannot be graphically renormalized.

So if we now perform a {\it full} renormalization group analysis on Eq.~(\ref{incompEOM}), {\it including graphical corrections}, the full recursion relations for the graphically unrenormalized parameters $\lambda$, $\gamma$, and $\Delta$ will be just what we obtained earlier from power counting; i.e.,
\begin{eqnarray}
\gamma^{\prime}&=&b^{z-\zeta}\gamma  ~~,
\label{gammarescale2}
\end{eqnarray}
\begin{eqnarray}
\Delta^{\prime}&=&b^{2(z-\chi)+1-d-\zeta}\Delta ~~,
\label{2Deltarescale}
\end{eqnarray}
 and 
\begin{eqnarray}
\lambda^{\prime}&=&b^{z+\chi-1}\lambda~~,
\label{2lambdarescale}
\end{eqnarray}
{\it exactly}. The other parameters ($D_T$ and ${c_0^2\over\rho}$) will get graphical renormalizations, so the values of the scaling exponents $z$, $\zeta$, and $\chi$ that will keep them fixed will no longer be the linear ones Eq.~(\ref{linexp}). We can, however, determine the exact values of $z$, $\zeta$, and $\chi$ that will give us a fixed point from the {\it exact} recursion relations Eqs.~(\ref{gammarescale2}--\ref{2lambdarescale}), which imply that, to get a fixed point, we must have
\beq
z-\zeta=0 ~ , ~ 2(z-\chi)+1-d-\zeta=0 ~ , ~ z+\chi-1=0 ~ .
\label{expcond}
\eeq
These three equations in three unknowns are easily solved for the exact values of these scaling exponents for $\gamma v_2<0$ for all spatial dimensions $d$ in the range $2<d<5$:
\begin{eqnarray}
z&=&{d+1\over 3}=\zeta  ~~,
\label{zexact1}
\end{eqnarray}
\begin{eqnarray}
\chi&=&{2-d\over 3}  ~~.
\label{chiexact1}
\end{eqnarray}
Note that, when $d$ equals  the critical dimension ($d=d_c=5$),  these match on to the values $z=\zeta=2$, $\chi={3-d\over 2}=-1$ of these exponents predicted by the linear theory, as they should.

The fact that $\chi<0$ for all $d$ in the range $2<d<5$ implies that velocity fluctuations get smaller as we go to longer and longer length scales; this implies the existence of long ranged order (i.e., a non-zero average velocity $\overline\bv\ne{\bf 0}$) in all of those spatial dimensions. The physically realistic case in this range is, of course, $d=3$. 

We can also calculate the scaling of  correlations of velocity fluctuations from these exponents. For example, the usual scaling arguments imply that 
\begin{eqnarray}
C_{vv}({\bf r})&\equiv&\overline{\delta{\bf v}({\bf r}+{\bf R},t)\cdot\delta{\bf v}({\bf R},t)}\nonumber\\&=& b^{2\chi}C_{vv}\left(b^{-1}\brp, b^{-\zeta}r_\parallel\right) ~ .
\label{vvrealscale}
\end{eqnarray}
Choosing $b=\rp/\chip$, where $\xi$ is some fixed microscopic length, this can be rewritten

\begin{eqnarray}
C_{vv}({\bf r})&=&
\rp^{2\chi} \chip^{-2\chi}C_{vv}\left(\chip, {\rpar\over\left({\rp\over\chip}\right)^\zeta}\right)\equiv\rp^{2\chi} g\left({\left({\rpar\over\chipar}\right)\over\left({\rp\over\chip}\right)^\zeta}\right)~ ,\nonumber\\\,\,\,\,\,\,\, 
\label{vvrealscale1}
\end{eqnarray}
where we've defined the scaling function
\begin{eqnarray}
g(x)&\equiv&
 \chip^{-2\chi}C_{vv}\left(\chip, x\chipar\right)~ ,\,\,\,\,\,\,\, 
\label{gdef}
\end{eqnarray}
with $\chipar$ another microscopic length that we've introduced to make the argument of $g$ dimensionless.

We can determine the limiting behaviors of $g$ by noting that we expect 
$C_{vv}({\bf r})$ to depend only on $\rpar$ for ${\rpar\over\chipar}\gg\left({\rp\over\chip}\right)^\zeta$; inspection of Eq.~(\ref{vvrealscale1}) reveals that this can only happen if
\begin{eqnarray}
g(x\gg1)\propto x^{\chi/\zeta}
~ ,\,\,\,\,\,\,\, 
\label{glarge}
\end{eqnarray}
because only then will the $\rp$ dependence drop out. 
This implies that 
\begin{eqnarray}
C_{vv}({\bf r})&\propto&
\rpar^{2\chi/\zeta}=\rpar^{2(2-d)\over d+1}=\rpar^{-1/2}\,\,\,\,\,, ~  \,\,\,\,
{\rpar\over\chipar}\gg\left({\rp\over\chip}\right)^{d+1\over3}~ ,\nonumber\\\,\,\,\,\,\,\, 
\label{vvrparscale}
\end{eqnarray}
where we have used our exact results Eqs. (\ref{chiexact1},\ref{zexact1}) for $\chi$ and $\zeta$, and the final equality holds in $d=3$.

Likewise, 
for ${\rpar\over\chipar}\ll\left({\rp\over\chip}\right)^\zeta$, we expect 
$C_{vv}({\bf r})$ to depend only on $\rp$; this implies
\begin{eqnarray}
g(x\to0)\to \rm{constant}\ne0
~ ,\,\,\,\,\,\,\, 
\label{gsmall}
\end{eqnarray}
which implies
\begin{eqnarray}
C_{vv}({\bf r})&\propto&
\rp^{2\chi}=\rp^{2(2-d)\over 3}=\rp^{-2/3}\,\,\,\,\,, ~  \,\,\,\,
{\rpar\over\chipar}\ll\left({\rp\over\chip}\right)^{d+1\over3}~ ,\nonumber\\\,\,\,\,\,\,\, 
\label{vvrperpscale}
\end{eqnarray}
where again the final equality holds in $d=3$.

Note that since $\chi={d+1\over3}>1$, for most directions of $\br$, ${\rpar\over\chipar}\ll\left({\rp\over\chip}\right)^{d+1\over3}$ for large $r$, so Eq.~(\ref{vvrperpscale}) will hold. It is only for $\br$ very nearly parallel to the direction of mean flock motion (that is, for  $\theta_r\ll1$) that the other limit Eq.~(\ref{vvrparscale}) will apply.
 
It is instructive, particularly for comparison with the $\gamma v_2>0$ case, to rewrite the scaling law Eq.~(\ref{vvrealscale1}) in polar coordinates: $\rp=r\sin\theta_r$, $\rpar=r\cos\theta_r$. This leads, in $d=3$, directly to equation Eq.~ (\ref{Creal1}), with the definition
\be
f_T(x)\equiv x^{-2/3} g(x^{-4/3}) ~ .
\label{fdef}
\ee
The limiting behaviors for large and small $\theta_r$ given on the last two lines
of Eq.~(\ref{Creal1}) follow directly from the limiting behaviors  Eq.~(\ref{glarge}) and  Eq.~(\ref{gsmall}) of the scaling function $g$, together with the relation Eq.~(\ref{fdef}) between $f_T$ and $g$.

We can also, by virtually identical reasoning, derive similar scaling laws for the correlations in Fourier space:
\begin{eqnarray}
\overline{|{\bf v}_{_\perp}({\bf q})|^2}&=&\qpar^{-2} h\left({\qpar
 \over
\mqp^\zeta }
\right)
\nonumber\\
&\propto&\left\{
\begin{array}{ll}
\mqp^{-2\zeta}\,\,\,\,\,\,\,\,\,,
&{\qpar\over\Lambda}\ll \left({\mqp\over\Lambda}\right)^\zeta\,,
\\ \\
\qpar^{-2}\,\,\,\,\,\,\,\,\,,
&{\qpar\over\Lambda}\gg \left({\mqp\over\Lambda}\right)^\zeta\,.
\\ 
\end{array}\right.
\label{vsc2}
\end{eqnarray}

Density fluctuations $\overline{|\delta\rho(\bq)|^2}$ in this case are unaffected by the non-linearities, since the parameters $\gamma$, $v_2$, $\Delta$, and $c_0$ that control them for all directions of $\bq$ are unrenormalized. (To see that these are indeed  the only parameters that  affect $\overline{|\delta\rho(\bq)|^2}$, look at Eq.~(\ref{rhoanglefluc}) for the case $\gamma v_2<0$.) Hence, our linear result, Eq.~\ref{rhoanglefluc}), applies for all directions of $\bq$, which implies 
\beq
\overline{|\delta\rho(\bq)|^2}\propto1/q^2
\eeq
for all directions of $\bq$. Fourier transforming this result back to real space implies
\begin{eqnarray}
C_{\rho\rho}({\bf r})=h_\rho(\theta_r)/r^{d-2} \ ,
\label{Crhorealgvt<0}
\end{eqnarray}
where $h_\rho(\theta_r)$ is a smooth, analytic $\cO(1)$ function of $\theta_r$. Comparing this result with the general form Eq.~(\ref{Crhoreal}), we see that in the notation of that equation, we have $\beta_\rho=0$ and $\chi_\rho=d-2$, as claimed in the introduction for this case.

\subsection{$\gamma v_2<0$, $d=2$}


In $d=2$, as we've already discussed, there is no transverse component $\vt$ of the velocity. As also discussed in the section on the linear theory, this implies that, in the $\gamma v_2<0$ case which is the topic of this section, there are {\it no} directions of $\bq$ in which the linear theory predicts a divergence of $\vq$ stronger than $1/q^2$ as $q\to0$, in contrast to the $d>2$ case, in which some of the components of the velocity - namely
$\vt$ - diverge (in the linear theory) like $1/q^4$ in certain directions (specifically, perpendicular to the mean velocity $\overline\bv$). This means that the power counting will be quite different in $d=2$, where these fluctuations are absent.

Before undertaking that revised power counting, however, we first recall, as discussed at the end of our treatment of the linearized theory, that even from that linearized theory, we can already see that non-linear effects {\it must} be important, and must invalidate, to some degree, the linear results. 
This is because the divergence of $\rsv$ predicted by the linearized theory implies that $\overline\bv={\bf 0}$ in this case. This means that either the state of the flock is isotropic, or non-linearities must stabilize long-ranged order. Either possibility implies that non-linearities {\it must} become important, because the linearized results for the velocity correlations Eq.~(\ref{vLanglefluc}) are anisotropic, but, at the same time, imply divergent fluctuations, which implies the system must be {\it isotropic} on sufficiently large length scales. This self-contradiction implies that the result Eq.~(\ref{vLanglefluc}) must be incorrect in $d=2$; we have just proved this by contradiction. But since that result depended {\it only} on linearizing the equations of motion, this must mean that the linear theory breaks down in $d=2$. 

We will now confirm this by power counting. We must keep $c_0$, $\gamma$, $v_2$, $\Delta$, and $\rho_0$ fixed. This implies
\begin{eqnarray}
\gamma^{\prime}&=&b^{z-\zeta}\gamma  ~~,
\label{gammarescale_d=2}
\end{eqnarray}
\begin{eqnarray}
v_2^{\prime}&=&b^{z-\zeta}v_2  ~~,
\label{v2scale_d=2}
\end{eqnarray}
\begin{eqnarray}
\left({c_0^2\over\rho_0}\right)^{\prime}&=&b^{{\chi_\rho-\chi+z-1}}\left({c_0^2\over\rho_0}\right)  ~~,
\label{c_0rescale_d=2}
\end{eqnarray}
\begin{eqnarray}
\Delta^{\prime}&=&b^{2(z-\chi)-1-\zeta}\Delta ~~.
\label{Deltarescale_d=2}
\end{eqnarray}
\begin{eqnarray}
\rho_0^{\prime}&=&b^{{\chi-\chi_\rho+z-1}}\rho_0  ~~,
\label{rho_0rescale_d=2}
\end{eqnarray}
where we have set $d=2$ throughout.

We can thus keep the scale of the fluctuations of $\rho$ and $v_L$ fixed by choosing the exponents $z$, $\zeta$, $\chi$, and $\chi_\rho$ to keep the above parameters fixed, which means the exponents must obey
\begin{eqnarray}
z-\zeta=0  ~~,
\label{gammafix1}
\end{eqnarray}
\begin{eqnarray}
\chi_\rho-\chi+z-1=0  ~~,
\label{c_0fix1}
\end{eqnarray}
\begin{eqnarray}
\chi-\chi_\rho+z-1=0  ~~,
 ~~,
\label{rhofix1}
\end{eqnarray}
\begin{eqnarray}
2(z-\chi)-1-\zeta=0 ~~.
\label{Deltafix1}
\end{eqnarray}
This system of linear equations can readily be solved for all of the exponents, yielding
\begin{eqnarray}
z_{\rm{lin}}=\zeta_{\rm{lin}}=1  ~,~\chi_{\rm{lin}}=\chi_{\rho,\rm{lin}}=0 ~~.
\label{linexp1}
\end{eqnarray}
The subscript ``lin" in these expressions denotes the fact that we have determined these exponents ignoring the effects of the non-linearities in the equations of motion Eqs.~(\ref{vEOMbroken}--\ref{cons broken}). We now use them to determine in which  non-linearities are important in $d=2$.

Upon the rescalings Eq.~(\ref{rescale}), the  non-linear terms
$\lambda$ and
$g_{1,2,3,4}$ in the $\vp$ equation of motion Eq.~(\ref{vEOMbroken})  obey

\begin{eqnarray}
\lambda^{\prime}&=&b^{z+\chi-1}\lambda= \lambda ~~,
\label{lambdarescale3}
\end{eqnarray}

\begin{eqnarray}
g_1^{\prime}&=&b^{z+\chi_\rho-\zeta}g_1 =g_1 ~~,
\label{g1rescale3}
\end{eqnarray}

\begin{eqnarray}
g_2^{\prime}&=&b^{z+\chi_\rho-\zeta}g_2 =g_2 ~~,
\label{g2rescale3}
\end{eqnarray}

\begin{eqnarray}
g_3^{\prime}&=&b^{\chi_\rho}g_3 =g_3 ~~.
\label{g3rescale3}
\end{eqnarray}

\begin{eqnarray}
g_4^{\prime}&=&b^{z+2\chi_\rho-\chi-1}g_4 =g_4 ~~.
\label{g4rescale3}
\end{eqnarray}
Doing the same for $\lambda_\rho$, and $w_{1,2}$ non-linearities in the $\rho$ equation of motion Eq.~(\ref{cons broken}) gives
\begin{eqnarray}
\lambda_\rho^{\prime}&=&b^{z+\chi-1}\lambda_\rho= \lambda_\rho ~~,
\label{lambdarhorescale}
\end{eqnarray}

\begin{eqnarray}
w_1^{\prime}&=&b^{z+\chi_\rho-\zeta}w_1 =w_1 ~~,
\label{w1rescale}
\end{eqnarray}

\begin{eqnarray}
w_2^{\prime}&=&b^{z+\chi-\zeta}w_2 =w_2 ~~,
\label{w2rescale}
\end{eqnarray}

We see that  all  eight of these non-linear couplings are marginal in this case. This implies that they will all give rise to logarithmic changes to the linear theory.

Because these changes are only logarithmic, they will only be apparent on literally astronomical length scales at small disorder strength. For example, $\gamma$ will presumably get corrections $\delta\gamma$ which behave like $\delta\gamma=\rm{constant}\times\Delta\ln(L/\xi)$, where $\xi$ is some microscopic length, $L$ is the spatial extent of the system, and the ``constant" is independent of the disorder strength $\Delta$. For these corrections to become comparable to the ``bare" $\gamma$, we clearly must go to system sizes
\beq
L\sim\xi\exp( {\rm{constant}\over\Delta})\equiv\xi\exp( {\Delta_c\over\Delta}) \ ,
\label{qnl}
\eeq
which is such a strong function of the disorder strength $\Delta$ that it can easily become astronomically large. For example, even if we take $\xi$ to be an interparticle distance, if we take a not particularly small value of $\Delta=\Delta_c/9$, we get $L\sim8000\xi$, which in two dimensions would mean a flock of $8000^2=64$ million flockers! So many simulated flocks will be too small to see the logarithmic effects just described,  and the linear theory described earlier should work.







We could have anticipated our result that the non-linearities are marginal by another line of reasoning. 
Specifically, our result Eq.~(\ref{vLanglefluc})  is anisotropic. But this   is logically inconsistent in two dimensions  with our prediction (\ref{vrflucscale2}) of diverging velocity fluctuations, because such diverging fluctuations imply that rotation invariance is {\it not} broken. If rotation invariance is not broken, correlation functions must be isotropic. This self-inconsistency of the linear theory in $d=2$ implies that the linear theory {\it must} be incorrect in $d=2$.
The marginal non-linearities just identified provide the mechanism for this breakdown of the linearized theory; the fact that they are just marginal reflects the fact that the fluctuations that restore isotropy only diverge logarithmically in $d=2$.

What happens once we get to big enough systems, or strong enough disorder, to see these logarithmic effects? To answer this question with certainty would require a dynamical renormalization group analysis incorporating all eight of the marginal non-linearities. We estimate that $8^3=512$ Feynmann graphs would have to be evaluated just to compute the renormalization of these non-linearities themselves; an additional $8^2=64$ graphs would have to be done for, at the very least, each of the noise strength $\Delta$ and the diffusion coefficient $D_T$. These $640$ Feynmann graphs would then lead to $10$ differential equations with a total of $640$ terms, which would then have to be analyzed to determine the ultimate scaling behavior of the system. This calculation is beyond our stamina, and we have not attempted it. 

Instead, we will engage in informed speculation about what the result of such an analysis would be. We suspect that the problem is like a  two-dimensional nematic \cite{2dnem, act2dnem}. In such a nematic, both equilibrium\cite{2dnem} and active\cite{act2dnem}, with unequal Frank constants, linear theory predicts anisotropic director correlations, and logarithmically diverging real space director fluctuations. As here, so there these two results are mutually inconsistent, since logarithmically diverging fluctuations will restore isotropy. In that 2d nematic problem, the paradox is resolved, and 
isotropy is restored, by  slow (logarithmic) renormalization of the Frank constants towards equality \cite{2dnem}. We suspect something similar happens here.

We also note that those subtle effects should only become apparent on length scales that grow like $\exp[\rm{constant}/\Delta]$ for small $\Delta$, which will become astronomically large length scales if the noise strength $\Delta$ is small, as it  is in our simulations. This appears to be the case in our simulations, since, as shown in Fig. ~\ref{Fig3Correlation}, they still exhibit considerable anisotropy.

Turning our attention now to density fluctuations, we see that, up to logarithmic corrections, 
density fluctuations $\overline{|\delta\rho(\bq)|^2}$ again obey
\beq
\overline{|\delta\rho(\bq)|^2}\propto1/q^2
\eeq
for all directions of $\bq$. Fourier transforming this result back to real space implies, as before, that
\begin{eqnarray}
C_{\rho\rho}({\bf r})=h_\rho(\theta_r)\times\ln{r} \ ,
\end{eqnarray}
where $h_\rho(\theta_r)$ is a smooth, analytic $\cO(1)$ function of $\theta_r$. Comparing this result with the general form (\ref{Crhoreal}), we see that in the notation of that equation, we have $\beta_\rho=0$ and $\chi_\rho=d-2=0$, as claimed in the introduction for this case.

\subsection{$\gamma v_2>0$, $d=2$}

For this case, there is no longer a transverse velocity $\vt$. However, because the Fourier transformed longitudinal field $v_L(\bq)$, and the density $\rho(\bq)$ now both exhibit anisotropic and divergent fluctuations with the same scaling as those exhibited by $\vt$ in higher dimensions, the eight vertices involving these all become as relevant as the $\lambda$ vertex is in the $\gvt<0$ case considered in subsection(A) above. Since the RG eigenvalue of that vertex is  ${5-d\over2}$ (see Eq.~(\ref{lambdarescale})), and since we are considering $d=2$ here, these vertices are strongly relevant, and will produce much stronger than logarithmically divergent corrections. Treating these, however, is fraught with the same difficulties arising from the large number of relevant non-linearities as just discussed in  the $\gamma v_2<0$, $d=2$ section, compounded by the fact that $d=2$ is so far below the upper critical dimension $d_c=5$ that a perturbation theory approach, even if practical, will not yield quantitatively reliable results.

However, our experience with the annealed noise problem suggests a way out. In that annealed case, the {\it assumption} that below the critical dimension only {\it two} of the non-linearities, namely the convective $\lambda$ and $\lambda_\rho$ terms in (\ref{vEOMbroken}) and (\ref{cons broken}), respectively, are actually relevant, simplifies the problem so much that it is possible to determine exact exponents in $d=2$.

This assumption may seem dubious, or, worse, contradictory with the power counting we've just done, which says that all eight non-linearities $\lambda$, $\lambda_\rho$, $g_{1,2,3,4}$, and $w_{1,2}$ become relevant in the same dimension. However, the power counting argument just given does not, in fact, rule out this possibility. This
is because the power counting argument was done at the {\it linearized} fixed point, at which all of  the non-linearities are zero. The relevance of these non-linearities $d\le d_c^{NL}=5$ therefore means that  for those dimensions, the system will flow away from this linear fixed point to a new, nonlinear  fixed point. What the values of the non-linearities are at that fixed point can only be determined by a full-blown renormalization group  analysis, which, as we've discussed above, is impossible in practice in the dimensions of physical interest. The only thing we know for sure is that {\it at least one} of the eight non-linearities $\lambda$, $\lambda_\rho$, $g_{1,2,3,4}$, and $w_{1,2}$  {\it must} be non-zero at the stable RG fixed point for $d<5$ (since we've just shown that the fixed point with {\it all} of them  zero- i.e., the linear fixed point-is unstable for $d<5$). So it is entirely possible (although obviously by no means guaranteed) that $\lambda$ and $\lambda_\rho$ {\it are} the only non-zero non-linearities at the new fixed point.

There are precedents for this (that is, for terms that appear relevant by simple power counting below some critical dimension $d_c$ actually proving to be irrelevant once "graphical corrections" -i.e., nonlinear fluctuation effects - are taken into account). One example  is the cubic symmetry breaking interaction \cite{Aharony} in the $O(n)$ model, which is relevant by power counting at the Gaussian fixed point for $d<4$, but proves to be irrelevant, for sufficiently small $n$, at the Wilson-Fisher fixed point that actually controls the transition for $d<4$, at least for $\epsilon\equiv4-d$ sufficiently small.

The assumption that, of all the potentially relevant non-linearities, only $\lambda\ne 0$ in the annealed problem, leads to exact exponents for that problem which agree extremely well with numerical simulations of flocking \cite{TT2,TT3,TT4,field}. Thus it seems that this assumption is, in fact, correct for the annealed problem, which gives us some hope that it might also work in the quenched disorder problem.

We now investigate the consequences that follow {\it if} $\lambda$ and $\lambda_\rho$ are the only relevant  nonlinearities. They are:

\noindent1) The changes in $\lambda$ and $\lambda_\rho$ coming from the  rescaling step of the dynamical RG are identical; that is, $\lambda$ and $\lambda_\rho$ have the same power counting.

\noindent2) The vertex $\lambda_\rho$ gets no graphical corrections because mass conservation is exact.

\noindent3) This implies that either the fixed point value $\lambda_\rho^*$ of $\lambda_\rho$ obeys $\lambda_\rho^*=0$, or  $\lambda$ gets no graphical corrections either (otherwise, $\lambda$ wouldn't be fixed).

\noindent4) There are no graphical corrections to $\lambda$ if $\lambda_\rho=\lambda$, because then the system exhibits pseudo-Galilean invariance. As discussed earlier, when this happens, there can be no graphical
renormalization of either $\lambda$ or $\lambda_\rho$.

\noindent5) Points 1) and 4) taken together  imply that if $\lambda\le\lambda_\rho$ initially, $\lambda$ remains $\le\lambda_\rho$ upon renormalization. 

We can now prove by contradiction that $\lambda_\rho^*\ne0$ at the fixed point. Let's assume the contrary; then

\noindent6) this implies that if $\lambda_\rho$ renormalizes to $0$ under the RG, $\lambda$ does so as well.

\noindent7) But we know by power counting at the linear fixed point that  such a  fixed point, with all non-linearities vanishing, is unstable (the only suh fixed point is the linear fixed point we discussed earlier). Therefore, the system cannot flow to such a fixed point. 

\noindent8) Therefore, we have just proved by contradiction that any system which starts with $\lambda\le\lambda_\rho$ (and there must be some such systems, because no symmetry forbids it) initially must flow to a fixed with $\lambda_\rho\ne0$.

\noindent9) Therefore,  by point 3) above, $\lambda$ gets no graphical corrections at the fixed point.

\noindent10) Since both the $\lambda$ and the $\lambda_\rho$ vertices are total derivatives in $d=2$, there is no graphical correction to the disorder strength $\Delta$ either.

\noindent11) Points 9) and 10) taken together imply that, for the purposes of a self-consistent perturbation theory treatment, we can treat $\lambda$, $\Delta$ as constants, rather than wavevector dependent quantities.

\noindent12) Finally, because both vertices only involve $\perp$ derivatives, only $D_\perp$ and $D_{\rho\perp}$ get any graphical renormalization.

We will now use these observations to derive exact scaling laws for $\gvt>0$, $d=2$. This case is particularly important for comparison with our two-dimensional simulations, since the system we simulate proves to have $\gvt>0$.

We will now analyze this problem using a self-consistent perturbation theory approach.


This approach proceeds by treating the non-linearities in the equations of motion Eqs.~(\ref{vEOMbroken}--\ref{cons broken})
as a small perturbation on the linear theory, and calculating perturbatively the corrections they introduce. As usual, these corrections to the two point correlations we've calculated can be summarized by replacing all of the parameters of the linearized theory (e.g., the diffusion constant $D_{_\perp}$) with ``renormalized'' wavevector dependent quantities (e.g., $D_{_\perp}({\bf q})$) in the linearized expressions for the two point correlation functions. As equally usual, the perturbative calculation of these renormalized parameters can be represented by Feynmann graphs. See, e.g., \cite{FNS} for details.

\begin{figure}
\includegraphics[trim={0cm 1.5cm 0cm 2cm}, clip=true,width=2.5in]{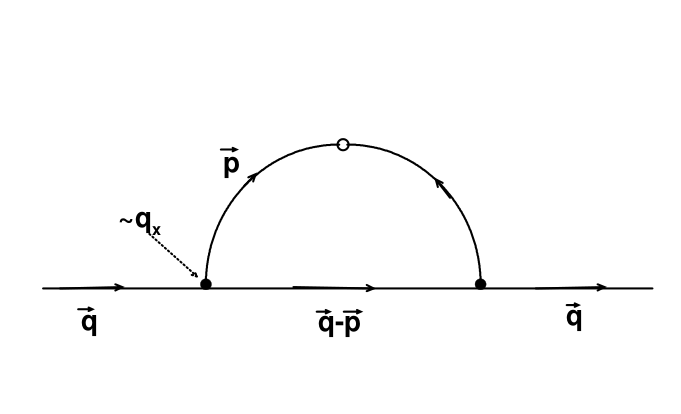}
 \caption{The Feynman diagram for a correction to the diffusion constant $D_{\perp}$.}
\label{diffusiondiagram}
\end{figure}

Consider, for example, the leading order graph illustrated in Fig.~\ref{diffusiondiagram}; this  leads to a correction to $D_\perp$ of the form:

\begin{equation}
\delta D_\perp({\bf q})=C\lambda^2 
\int{d^dp\over(2\pi)^d} \,\left\{\overline{|v_L({\bf p})|^2}G_L({\bf q}-{\bf p})
\right\},
\label{Dgraph1}
\end{equation}
where $C$ is an $\cO(1)$ constant, the exact value of which we will not need. In the harmonic approximation for $\overline{|v_L({\bf p})|^2}$ and $G_L({\bf q}-{\bf p})$ used earlier, the integral on the right hand side diverges in the infra-
red (small ${\bf p}$) limit, for ${\bf q}={\bf 0}$, for spatial dimensions $d\le5$. Of course, we are only interested in $d=2$ here; for $d>2$, the structure of the problem changes significantly, not least because of the presence of transverse components $\vt$ of $\vp$. Equation (\ref{Dgraph1}) should therefore be thought of as a generalization of the correction to $\Dp$ in $d=2$ to higher dimensions, {\it not} as an expression that's actually valid for a flock in those higher dimensions. We are simply doing this extension here to illustrate a point, which will become clear in a moment.



Writing the correction to $\Dp$ explicitly by using the propagators and correlation functions (\ref{rhoanglefluc}) and (\ref{vLanglefluc}) we obtained in the linear theory, but replacing the bare $D_\perp$ with $D_\perp(\bq)$ and the bare $D_{\rho\perp}$ by $\Dpr$, to be determined self-consistently. Doing so leads to: 
\begin{widetext}
\begin{eqnarray}
\delta D_\perp=C \lambda^2\int_{p_s>q_{IR}}{d^dp\over(2\pi)^d}
{v_2^3p_\parallel^4\Delta\Xi(\bq-{\bf p})\over \left[\left[c_0^2p_\perp^2-\gamma v_2p_\parallel^2\right]^2+\Xi({\bf p})^2p_\parallel^2\right]\left[\left[c_0^2|{\bf p}_\perp-\bq|^2-\gamma v_2(p_\parallel-q_\parallel)^2\right]^2+\Xi(\bq-{\bf p})^2(p_\parallel-q_\parallel)^2\right]}~,
\nonumber\\
\label{delD1}
\end{eqnarray}
\end{widetext}

The leading order graphical correction to $D_{\rho\perp}$ is identical to this, but with $\lambda$ replaced by $\lambda_\rho$. Since, as we argued above, our assumption that $\lambda$
and
$\lambda_\rho$ are the only relevant non-linearities implies that there are no graphical corrections to $\lambda$
and
$\lambda_\rho$. That assumption therefore also implies that the renormalization 
$\delta D_{\rho\perp}$ of $D_{\rho\perp}(\bq)$ is proportional to that $\delta D_\perp$ of $\Dp$.
Since these corrections prove to dominate the bare values, the bottom line is that $\Dpr\propto\Dp$. We will use this fact below to get a closed self-consistent equation for $\Dp$, whose solution will then, of course, determine {\it both} $\Dp$ and $\Dpr$.

As just noted, 
for small $\bq$, the ostensibly  ``small" correction (\ref{delD1}) actually dominates the bare value of $D_\perp$. We can therefore, for $d\le5$, replace $\delta D_\perp$ with the wavevector dependent, renormalized $D_\perp({\bf q})$ on the left hand side of Eq.~(\ref{Dgraph1}). 
Furthermore, since $D_\perp$ and $\Dpr$ are the {\it only} diffusion coefficients that diverge (if our assumption  that $\lambda$ and $\lambda_\rho$ are the only relevant non-linearities at the fixed point is correct), which  they dominate all of the other diffusion constants in the expression 
(\ref{Xidef}) for  the damping coefficient $\Xi({\bf p})$,

If we also replace the $D_s$'s   that appear implicitly on the right hand side (inside $\overline{|v_L({\bf p})|^2}$,$ \overline{|{\bf v}_T({\bf p})|^2}$,
$G_T({\bf q}-{\bf p})$, and $G_L({\bf q}-{\bf p})$) with $C^{\prime\prime}D_\perp({\bf q}-{\bf p})$ and $D_\perp({\bf p})$, where appropriate, we thereby make Eq.~(\ref{Dgraph1}) into a self-consistent integral equation for $D_\perp({\bf q})$.

This integral equation can be simplified by noting that,
as with the integrals for the real space fluctuations (Eq.~(\ref{vrfluc})), this integral is also dominated by wavevectors $\vec{p}$
whose direction $\theta_{\vec {p}}$ is close to the critical angle $\theta_c$ defined earlier.  We can therefore make the same
approximations for $\theta_{\vec {p}}$ near $\theta_c$ that we made earlier. We can simplify even further by noting
 that the scaling of the rather complicated integral on the right hand side of Eq.~(\ref{Dgraph1}) is the same as the scaling of  the
 same integral with $\vec{q}$ set $=\vec{0}$ in the integrand, but with infrared cutoffs of $p>q$ and $\delta\theta_{\vec{p}}>\delta\theta_
{\vec{q}}$ applied to the range of integration itself.

With these simplifications, Eq.~(\ref{delD1}) becomes a self-consistent equation for $D_\perp(\vec{q})$ :
\begin{widetext}
\begin{eqnarray}
D_\perp(\vec{q})={C\lambda^2\Delta v_2^3\sin^2\theta_c\cos^4\theta_c\over(2\pi)^2}\int_{q}^{\infty}p^{d-3}dp
\int_{\delta\theta_{\vec{q}}}^{\infty}{d\delta\theta_{\vec{p}}}{\left[v_2D_\perp(p, \delta\theta_{\vec{p}})+\gamma D_{\rho\perp}(p, \delta\theta_{\vec{p}})\right]\over \left({4\gamma v_2\over c_0^2} \delta\theta_{\vec{p}}^2+ [v_2D_\perp(p, \delta\theta_{\vec{p}})+\gamma D_{\rho\perp}(p, \delta\theta_{\vec{p}})]^2 p^2\sin^4\theta_c\cos^2\theta_c\right)^2}
\nonumber\\
\label{Dsc<}
\end{eqnarray}
\end{widetext}

We can solve this integral equation with the simple scaling ansatz:
\begin{eqnarray}
D_\perp(\vec{q})=q^{-\eta} g_\theta\left({\delta\theta_{\vec{q}}
 \over
q ^\alpha }
\right) \quad .
\label{Dsc<1}
\end{eqnarray}
which, due to the proportionality of $\Dp$ and $\Dpr$ noted earlier, implies
\begin{eqnarray}
\Dpr=\gimel q^{-\eta} g_\theta\left({\delta\theta_{\vec{q}}
 \over
q ^\alpha }
\right) \quad ,
\label{Dsc<2}
\end{eqnarray}
where $\gimel$ is an $\cO(1)$ constant.

Inserting this ansatz on {\it both} sides of Eq.~(\ref{Dsc<1}), and noting that this ansatz implies that, as $p\rightarrow 0$, $D_\perp(p, \delta\theta_{\vec{p}})>>D_\parallel$, the self consistent equation becomes:
\begin{widetext}
\begin{eqnarray}
q^{-\eta} g_\theta\left({\delta\theta_{\vec{q}}\over q ^\alpha}\right)
\propto \int_{q}^{\infty}p^{d-3}dp
\int_{\delta\theta_{\vec{q}}}^{\infty}
d\delta\theta_{\vec{p}}
{p^{-\eta} g_\theta\left({\delta\theta_{\vec{p}}
\over p^\alpha}\right)
\over \left(S \delta\theta_{\vec{p}}^2+p^2[p^{-\eta} g_\theta\left({\delta\theta_{\vec{p}}\over p^\alpha}\right)]^2\right)^2}
\label{Dsc<3}
\end{eqnarray}
\end{widetext}
where $S$ is an unimportant constant. Changing variables of integration from $p$ and $\delta\theta_{\vec{p}}$
to $P$ and $\Phi$ defined via:
$p\equiv qP$ and $\delta\theta_{\vec{p}}\equiv\Phi\delta\theta_{\vec{q}}$, we can pull all of the dependence on $q$ on the right hand side out in front of the integral, obtaining:

\begin{gather}
\begin{gathered}
q^{-\eta} g_\theta\left({\delta\theta_{\vec{q}}\over q ^\alpha}\right)
\propto q^{d-2-\eta}\delta\theta_{\vec{q}}\int_{1}^{\infty}P^{d-3-\eta}dP
\int_{1}^{\infty}
d\Phi\\
{g_\theta\left({\delta\theta_{\vec{q}}
\over q^\alpha}{\Phi\over P^\alpha}\right)
\over \left(S \delta\theta_{\vec{q}}^2\Phi^2+q^{2-2\eta}P^{2-2\eta} g_\theta^2\left({\delta\theta_{\vec{q}}
\over q^\alpha}{\Phi\over P^\alpha}\right)\right)^2} \\
\\
\propto q^{d-2-\eta+4\eta-4}\delta\theta_{\vec{q}}\int_{1}^{\infty}P^{d-3-\eta}dP
\int_{1}^{\infty}
d\Phi\\
{g_\theta\left({\delta\theta_{\vec{q}}
\over q^\alpha}{\Phi\over P^\alpha}\right)
\over \left(S \delta\theta_{\vec{q}}^2q^{2(\eta-1)}\Phi^2+P^{2(1-\eta)} g_\theta^2\left({\delta\theta_{\vec{q}}
\over q^\alpha}{\Phi\over P^\alpha}\right)\right)^2}
\label{Dsc3}
\end{gathered}
\end{gather}

Now, everything in the double  integral  on the right hand side of this expression is explicitly a function of the scaling combination
${\delta\theta_{\vec{q}}\over q^\alpha}$ except for the factor in the $S$ term in the denominator
proportional to ${\delta\theta^2_{\vec{q}}\over q^{2(1-\eta)}}$. To make the scaling ansatz work, we
must force this term to also be a function only of the scaling combination
${\delta\theta_{\vec{q}}\over q^\alpha}$; this can clearly be done by choosing:
\begin{equation}
\alpha=1-\eta~.
\label{alpha}
\end{equation}
Using this in Eq.~(\ref{Dsc3}) makes the double integral a scaling function of
${\delta\theta_{\vec{q}}\over q^\alpha}$; we call this function $g_I\left({\delta\theta_{\vec{q}}\over q^\alpha}\right)$. Then Eq.~(\ref{Dsc3}) reads:
\begin{gather}
\begin{gathered}
q^{-\eta} g_\theta\left({\delta\theta_{\vec{q}}\over q ^\alpha}\right)
\propto q^{d-6+3\eta}\delta\theta_{\vec{q}}g_I\left({\delta\theta_{\vec{q}}\over q^\alpha}\right)\propto \\ q^{d-6+3\eta+\alpha}\left({\delta\theta_{\vec{q}}\over q^\alpha}\right)g_I\left({\delta\theta_{\vec{q}}\over q^\alpha}\right)~.
\end{gathered}
\end{gather}
Now everything on the right is explicitly a function  of the scaling combination
${\delta\theta_{\vec{q}}\over q^\alpha}$ times a power of $q$, as is the left hand side. Hence, our ansatz works provided only that these two powers of $q$ are equal; this implies
\begin{equation}
-\eta=d-6+3\eta+\alpha ~.
\label{eta}
\end{equation}
The equations (\ref{alpha})  and (\ref{eta})  are two simple linear equations for the  scaling exponents
$\eta$ and $\alpha$; their solution is:
\begin{equation}
\eta={5-d\over3}~~,~~\alpha={d-2\over3}~.
\label{etaalpha}
\end{equation}

Recalling that the case of physical interest here is $d=2$, we see that $\eta=1$ and $\alpha=0$. The last result implies that in $d=2$, the anisotropy of scaling of the behavior of the diffusion constants $\Dp$ and $\Dpr$ is gone: the range $\delta\theta$ over which these vary is independent of $q$. 

The result $\eta=1$ implies the same is true of the correlation functions  
$\overline{|v_L(\bq)|^2}$ and $\overline{|\rho(\bq)|^2}$. This is most easily seen from equations (\ref{vLexp}) and (\ref{gLdef}): inserting the ans\"atze (\ref{Dsc<1})
and (\ref{Dsc<2}) into those equations, we see that both correlation functions
$\overline{|v_L(\bq)|^2}$ and $\overline{|\rho(\bq)|^2}$ are proportional to $1/q^2$ times a function of $\theta_q$ alone, because the coefficient $\aleph\propto1/q$. 
This means the divergence of $\overline{|v_L(\bq)|^2}$ and $\overline{|\rho(\bq)|^2}$ as $\theta_q\to\theta_c$ has been cut off by the divergence of $\Dp$ and $\Dpr$. Another way to say this is that both correlation functions $\overline{|v_L(\bq)|^2}$ and $\overline{|\rho(\bq)|^2}$ now scale like $1/q^2$ for {\it all} directions of $\bq$, even  $\theta_q=\theta_c$.

A remnant of the divergence of these correlations at $\theta_q=\theta_c$ predicted by the linear theory persists, however. This is because the cutoff of that divergence is caused by the $1/q$ divergence of $\Dp$ and $\Dpr$ that we've just found. Since that divergence arises from fluctuations induced by the disorder, it follows that at small disorder, the {\it coefficient}
of that $1/q$ divergence (which is non-universal, unlike the exponents $\alpha$ and $\eta$, which are universal) will be small. Hence, the correlation functions $\overline{|v_L(\bq)|^2}$ and $\overline{|\rho(\bq)|^2}$ will have very large peaks at $\theta_q=\theta_c$ when plotted versus $\theta_q$ at fixed $q$. Indeed, putting in $\Dp$ and $\Dpr$ diverging like $1/q$ with small coefficients into equations (\ref{vLexp}) and (\ref{gLdef}), and noting that when those coefficients are small, these terms only matter near $\theta_q=\theta_c$ implies that, for all $\theta_q$, the correlation functions 
$\overline{|v_L(\bq)|^2}$ and $\overline{|\rho(\bq)|^2}$ can be well approximated by
\begin{equation}
q^2 \overline{ |v_{\perp}({\bf q})|^2 } \propto \frac{\Delta  \cos^2(\theta)}{(\sin^2(\theta)-\tan^2(\theta_c) \cos^2(\theta))^2+\delta} \ ,
 \label{vq2}
\end{equation}
and
\begin{equation}
q^2 \overline{ |\rho({\bf q})|^2 } \propto \frac{\Delta  \sin^2(\theta)}{(\sin^2(\theta)-\tan^2(\theta_c) \cos^2(\theta))^2+\delta} \ ,
 \label{rhoq2}
\end{equation}
with $\delta$ small.

We will see later that Eq.~(\ref{vq2}) fits our simulation data for  $\overline{ |v_{\perp}({\bf q})|^2 }$
extremely well (for a preview, see Fig. \ref{Fig3Correlation}), thereby supporting both the theory of this section, and the assertion that our simulated system has $\gvt>0$. Note that since this is true, the {\it linear} theory {\it incorrectly} predicts an actual divergence of the height of the peaks as $\bq\to {\bf 0)}$; the fact that this divergence is in fact cutoff shows the importance of the nonlinear corrections. It is also those corrections, and their suppression of this divergence, that make quasi-long-ranged order possible in these systems. 

All our earlier comments about restoration of full isotropy via further log corrections apply to this case as well. Our simulations are clearly of too small a system to be in this regime. As noted earlier, this is not surprising, due to the exponential divergence of the length scale for crossover to complete isotropy eqn. (\ref{qnl}). as $\delta\to0$.

Turning our attention now to density fluctuations, we see that, at least of our conjecture that $\lambda$ and $\lambda_\rho$ are the only relevant non-linearities is correct, this is very much like the $\gvt<0$, $d=2$ case. In particular, up to logarithmic corrections, 
density fluctuations $\overline{|\delta\rho(\bq)|^2}$ again obey
\beq
\overline{|\delta\rho(\bq)|^2}\propto1/q^2
\eeq
for all directions of $\bq$. Fourier transforming this result back to real space implies, as before, that
\begin{eqnarray}
C_{\rho\rho}({\bf r})=h_\rho(\theta_r)\times\ln{r} \ ,
\label{Crhorealgvt<0d=2}
\end{eqnarray}
where $h_\rho(\theta_r)$ is a smooth, analytic $\cO(1)$ function of $\theta_r$. Comparing this result with the general form (\ref{Crhoreal}), we see that in the notation of that equation, we have $\beta_\rho=0$ and $\chi_\rho=d-2=0$, as claimed in the introduction for this case.

\subsection{$\gamma v_2>0$, $d>2$}

This is the most complicated of our four cases, and the one about which we know the least. One thing we {\it do} know with certainty is that there {\it will} be anomalous hydrodynamics in these systems for all 
spatial dimensions $d\le5$, since {\it all} of the fields $\vt$, $v_L$, and $\rho$ exhibit anisotropic, strongly diverging fluctuations in the linearized approximation; indeed, in that linearized approximation they all have the same divergences with as $q\to0$, and the same anisotropy in those divergences ($q^2$ for  
most directions of $\bq$, $q^4$ for $\bq$ for certain values of $\theta_q$), as the 
field $\vt$ does in the $\gamma v_2<0$, $d>2$. In addition, the ``phase space" associated with the regions of $\bq$ that show the strogner divergence is the same in both cases: a $d-1$ diemnsional subspace of the $d$-dimensional $\bq$ space. Of course, this subspace is a hypercone $\theta_q=\theta_c={\sqrt{\gvt}\over c_0}$ for the fields $\rho$ and $v_L$, while it is a plane (the $\bqp$ plane) for $\vt$, but from a power counting standpoint, this distinction is unimportant.

Therefore, we expect all eight of the non-linearities $\lambda$, $\lambda_\rho$, $g_{1,2,3,4}$, and $w_{1,2}$ to be relevant.
Treating these, however, leads us to  the same difficulties we encountered in the $\gamma v_2>0$, $d=2$ case just discussed, arising from the large number of relevant non-linearities. We also again face the difficulty that $d=3$ is so far below the upper critical dimension $d_c=5$ that a perturbation theory approach, even if practical, will not yield quantitatively reliable results.

In fact, things are even worse in this case. This is because, even if we assume that $\lambda$ and $\lambda_\rho$ are the only relevant vertices, as we did in the $\gamma v_2>0$, $d=2$
section, we still cannot get exact exponents, because the $\lambda$ vertex cannot be written as a total derivative. Recall that our argument for it being so writeable depended, in the $\gamma v_2<0$, $d>2$ 
case, on the effective divergencelessness of the velocity in that case. Here, we can not make that argument; instead, as just explained, $v_L$ and $\vt$ have fluctuations of  the same size in a scaling sense. Nor can we use the argument we made for the $\gamma v_2<0$, $d=2$ case, for which we argued that the $\lambda$ vertex was a total derivative because $\vp$ had only one component; here, it has $d-2>1$ components. 

The upshot of all of this is that we have no way to determine the exact scaling exponents in this case, even if we are willing to make some unverifiable conjectures about the structure of the RG flows. A few things are clear, however:

\noindent1) The scaling laws {\it will} be anomalous, for $d\le5$, which, obviously, includes the physically interesting case $d=3$.

\noindent2) This anomaly should make the fluctuations in the velocity and the density smaller than those predicted by the linear theory. This assertion is based partly on experience - this is what happens for flocks with annealed disorder \fl, and
in the other three cases we have just treated for flocks with quenched disorder - and partly on physical intuition. Specifically, the microscopic mechanism for the non-linear suppression of fluctuations in all the cases just described is the enhanced exchange of information brought about by the motion of the flockers. This is why the diffusion constants are renormalized upwards. The phenomenon is quite similar to turbulent mixing \cite{turbulence}. Clearly, this mechanism is just as active - indeed, more active - in flocks with quenched disorder.

\noindent3) Since the linear theory predicts that long-ranged order is only marginally - i.e., logarithmically - destroyed in $d=3$, point two implies that order should be better  in the full non-linear theory. Hence, it must be long-ranged; that is, we must have $\overline{\bv}\ne{\bf 0}$.

\noindent4) Finally, based on the structure of the linear theory, we expect the scaling structure of fluctuations in Fourier space to be the the same for $\theta_q$ near $\theta_c$ and $\theta_q$ near $\pi/2$. This implies that fluctuations in real space should have the same scaling structure for $\theta_r$ near $\pi/2-\theta_c$ and $\theta_r$ near $0$.
This in turn implies that 
the connected velocity autocorrelation function defined above  in $d>2$ 
is given by
\begin{eqnarray}
C_{vv}({\bf r})=C_L({\bf r})+C_T({\bf r})\,,
\end{eqnarray}
where $C_L({\bf r})$ and $C_T({\bf r})$   represent the contributions to $C_{vv}({\bf r})$ coming from $v_L$ and $\vt$fluctuations, 
and respectively
obey the scaling laws 
\begin{eqnarray}
C_L({\bf r})&=&r^{-{\Omega}} f_L\left(\delta\theta_r r^{\beta } 
\right) h_L(\theta_r)\nonumber\\&\propto&\left\{
\begin{array}{ll}
(\delta\theta_r r)^{2\chi}\,\,\,\,\,\,\,\,\,,
&{\delta\theta_r}\gg r^{-{\beta }}\,,
\\ \\
r^{-{\Omega}}\,\,\,\,\,\,\,\,\,,
&{\delta\theta_r}\ll r^{-\beta}\,,
\end{array}\right.
\label{CLreal2}
\end{eqnarray}
and
\begin{eqnarray}
C_T({\bf r})&=&r^{-\Omega} f_T\left(\theta_r r^{\beta } 
\right) \nonumber\\&\propto&\left\{
\begin{array}{ll}
(\theta_r r)^{2\chi}\,\,\,\,\,\,\,\,\,,
&{\theta_r}\gg r^{-\beta}\,,
\\ \\
r^{-\Omega}\,\,\,\,\,\,\,\,\,,
&{\theta_r}\ll r^{-\beta}\,.
\\ 

\end{array}\right.
\label{CTrealgen2}
\end{eqnarray}

In (\ref{CLreal}),  we've defined
$\delta\theta_r\equiv\theta_r+\theta_c-{\pi\over2}$. The exponents $\beta$ and $\Omega$ in (\ref{CLreal}) and (\ref{CTrealgen}) are determined by the other two unknown, but universal, exponents - the anisotropy exponents $\zeta$, and the roughness exponent $\chi$ -  via the relations
\begin{equation}
\beta=1-{1\over\zeta}\,\,,\,\,
\Omega=-2{\chi\over\zeta}\,.
\end{equation}

Note that $C_L({\bf r})$ and $C_T({\bf r})$ exhibit their strongest anisotropies in different directions: $C_L({\bf r})$ is most strongly anisotropic near $\theta_r={\pi\over2}-\theta_c$, while  $C_T({\bf r})$ is most strongly anisotropic near $\theta_r=0$. Thus, the full correlation function $C_{vv}({\bf r})$ exhibits strong anisotropy near both directions of ${\bf r}$.

Up to factors of $\rho_0$ and $v_2$, and a factor of $\tan^2\theta_q$,  the Fourier transformed density correlations $\overline{|\delta\rho({\bf q})|^2}$ are equal to those of  $\overline{|v_L({\bf q})|^2}$, as can be seen by comparing (\ref{rhoanglefluc}) and (\ref{vLanglefluc}). Since 
$\rho_0$ and $v_2$ are not divergently renormalized, they can be replaced by constants. This means that $\overline{|\delta\rho({\bf q})|^2}$ scales in exactly the same way with $\bq$ as $\overline{|v_L({\bf q})|^2}$. Fourier transforming back to real space, this implies that $C_{\rho\rho}({\bf r})$ scales exactly like $C_L(\br)$; that is 
\begin{eqnarray}
C_{\rho\rho}({\bf r})&=&r^{-{\Omega}} f_L\left(\delta\theta_r r^{\beta } 
\right) h_\rho(\theta_r)\nonumber\\&\propto&\left\{
\begin{array}{ll}
(\delta\theta_r r)^{2\chi}\,\,\,\,\,\,\,\,\,,
&{\delta\theta_r}\gg r^{-{\beta }}\,,
\\ \\
r^{-{\Omega}}\,\,\,\,\,\,\,\,\,,
&{\delta\theta_r}\ll r^{-\beta}\ .
\end{array}\right.
\label{Crhoreald=3}
\end{eqnarray}
 Comparing this result with the general form (\ref{Crhoreal}), we see that in the notation of that equation, we have $\beta_\rho=\beta$ and $\chi_\rho=\chi$, as claimed in the introduction for this case.

While we can say nothing definite in $d=3$ for the case 
$\gamma v_2>0$, it is tempting to 
conjecture that the exponents  $\zeta$ and $\chi$ take on the same values as for $\gamma v_2<0$ in $d=3$, which are $\zeta=4/3$, $\chi=-1/3$. If this is the case, then we obtain $\beta=1/4$ and $\Omega=1/2$. We really have no justification for this conjecture, however, other than the fact that an analogous conjecture for flocks with {\it annealed} disorder appears empirically to get the correct exponents for $d=3$.

\section{Giant number fluctuations}\label{Giant number fluctuations}

The most natural 
quantity to look at when studying density fluctuations is the
fluctuations of the number of particles is an imaginary ``box'' of some
volume $V_{\rm box}$ inside a flock of volume $V_{\rm flock} \gg V_{\rm
box}$.  We will take our ``box'' to be a $d$-dimensional hypercube of  side $L$
(e.g., an $L \times L$ square in $d = 2$, or an $L \times L \times L$
cube in $d = 3$). The mean squared number fluctuations $\overline{\delta N^2} 
\equiv \left< N^2\right> - \left< N\right>^2$ can readily be related to
the real space correlations $C_{\rho\rho} (\br)$:
\begin{eqnarray}
\overline{\delta N^2} &=&\int_V d^dr d^dr^{\prime}\overline{\delta  \rho
(\br)\delta \rho (\br^{\prime})}\\ \nonumber &=&\int_{V}d^dr
d^dr^{\prime} \,C_{\rho\rho}\left( \br - \br^{\prime}\right)
\label{delN1}
\end{eqnarray}
where the subscript $V$ denotes that the integrals are over $\br$ 
and $\br^{\prime}$'s contained within our experimental ``box''. 

For three of our four cases,  namely, $d=2$, for both signs of $\gvt$, and $d=3$, $\gvt<0$, $C_{\rho\rho}({\bf r})$ is proportional to $r^{2-d}$ for all directions $\theta_r$ of $\br$.  Using this in (\ref{delN2}) gives
\begin{eqnarray}
\overline{\delta N^2} =\int_V d^dr \,d^dr^{\prime}|\br - \br^{\prime}|^{2-d}g_\rho(\theta_{\br-\br^{\prime}}) \ .
\label{delN3cases}
\end{eqnarray}
Making the changes of variables $\br \equiv
\bR L, \br^{\prime} \equiv \bR^{\prime}L$, we obtain
\begin{eqnarray}
\left<\delta N^2 \right> = L^{2+d} \int_{V_1}d^dR \,
d^dR^{\prime}  |\bR - \bR^{\prime}\: |^{ 2-d}  g_\rho
\left(\theta_{\bR - \bR^{\prime}} \right) \ .
\nonumber\\
\label{delN3cases2}
\end{eqnarray}
where $V_1$ denotes that the integrals are over $\vec{R}$ and 
$\vec{R}^{\prime}$ contained in a {\it unit} hypercube.  Hence, this
integral has {\it no} dependence on $L$.  Therefore (\ref{delN3cases2}) implies
\begin{eqnarray}
\left<\delta N^2 \right> = L^{2+d}  \times {\rm constant}
\label{delN3cases3}
\end{eqnarray}
where the constant is independent of $L$.  This can be rewritten in 
terms of the mean number $\left<N\right>$ of critters in the box, using
the fact that the average density $\rho _0$ is well-defined. Hence,
$\left<N\right> = \rho _0 L^d$, or $L = \left({\left<N\right> \over\rho
_0 } \right)^{1 \over d}$.  Using this in (\ref{delN4}) and taking the
square root of both sides gives:
\begin{eqnarray}
\sqrt {<\delta N^2>} \sim <N>^{\phi(d)}
\label{Nfluc0}
\end{eqnarray}
with
\begin{eqnarray}
\phi(d)=\frac{1}{2}+\frac{1}{d}   \ .
\label{phi3cases}
\end{eqnarray}

For the remaining case $d=3$, $\gvt>0$, $C_{\rho\rho}(\br)$ is given by (\ref{Crhoreal}) with $\beta_\rho=\beta=1-{1\over\zeta}$,  $\chi_\rho=\chi$, and $\Omega_\rho=\Omega=-2{\chi\over\zeta}$.

Using
this for $C_{\rho\rho}\left( \br -
\br^{\prime}\right)$ in (\ref{delN1}) gives
\begin{eqnarray}
\overline{\delta N^2}  &=&\int_V d^dr d^dr^{\prime}   |\br- 
\br^{\prime}|^{2\chi\over\zeta} g_\rho \left(\delta\theta_{\br -
\br^{\prime}}|\br -
\br^{\prime}|^{\beta} \right) \ .
\nonumber\\
\label{delN2}
\end{eqnarray}

Changing variables of integration from $\br^{\prime}$ to 
${\bf R}\equiv \br-\br^{\prime}$ gives
\begin{eqnarray}
\overline{\delta N^2}  &=&\int_V d^dr \int_Vd^dR  \, R^{2\chi\over\zeta} g_\rho \left(\delta\theta_{{\bf R}}R^{\beta} \right) \ .
\nonumber\\
\label{delN3}
\end{eqnarray}
Since the integrand is now independent of $\br$, we  the $\br$ integral trivially gives the volume $L^d$ of our box, so we have 
\begin{eqnarray}
\overline{\delta N^2}  &=&L^d \int_Vd^dR  \, R^{2\chi\over\zeta} g_\rho \left(\delta\theta_{{\bf R}}R^{\beta} \right) \ .
\nonumber\\
\label{delN4}
\end{eqnarray}
Now let's evaluate the integral over $\bR$ in this expression in hyperspherical coordinates. Since the integrand only depends on one of the polar angle $\theta_R$, the integrals 
over the remaining $d-2$ azimuthal angles just give a factor of $S_{d-1}$, defined as the surface area of a $d-1$-dimensional sphere of unit radius. Doing those integrals therefore leaves us with
\begin{widetext}
\begin{eqnarray}
\overline{\delta N^2}  &=&S_{d-1}L^d \int_0^L dR  \, R^{d-1}\int d\theta_R \sin^{d-1}(\theta_R)\,R^{2\chi\over\zeta}g_\rho \left(\delta\theta_{{\bf R}}R^{\beta} \right) \ .
\label{delN5}
\end{eqnarray}
\end{widetext}
The alert reader will note that, strictly speaking, this equation is not correct, since the range of integration of the magnitude $R$ of $\bR$ is not always $L$, but depends on the direction of $\bR$, since our cubic box is not spherically symmetric. However, the scaling  with $L$ of the correct integral will quite clearly be the same as that of the above integral, since the extent of the box along {\it any} direction is of order $L$.

Now let's split the integral $\int d\theta_R$ into two regions: one for small $\delta\theta$, specifically  covering the regime $|\delta\theta_R|\lesssim R^{-\beta}$; the other covering the regime (actually two regimes, one for positive, and one for negative, $\delta\theta_R$) covering $|\delta\theta_R|\gg R^{-\beta}$. Rather unimaginatively calling the integral over the small $\delta\theta$ regime $I_<$, and that over the large $\delta\theta$ regime $I_>$, and using the limiting forms from equation (\ref{Crhoreal}), we see that 
\beq
I_>\propto R^{2\chi} \ ,
\label{I>}
\eeq
while 
\beq
I_<\propto R^{-{2\chi\over\zeta}-\beta} \ .
\label{I<}
\eeq
Hence the ratio $I_</I_>\propto R^{2\chi(1-{1\over\zeta})-\beta}=R^{(1-{1\over\zeta})(2\chi-1)}$. Since $\zeta>1$ and $\chi<0$, we see that the exponent in this expression is $<0$, which means that for large $R$, $I_>\gg I_<$. Therefore dropping $I_<$, and using (\ref{I>}) as a good approximation to the full angular integral in (\ref{delN5}), we obtain
\begin{eqnarray}
\overline{\delta N^2}  &\propto&L^d \int_0^L dR  \, R^{2\chi+d-1} \ .
\label{delN6}
\end{eqnarray}
Changing variables of integration from $R$ to $u\equiv{R\over L}$ gives
\begin{eqnarray}
\overline{\delta N^2}  &\propto&L^{2(\chi+d)} \int_0^1 du  \, u^{2\chi+d-1} \ .
\label{delN7}
\end{eqnarray}
As before, the
integral in this expression has {\it no} dependence on $L$.  Therefore (\ref{delN3cases2}) implies
\begin{eqnarray}
\left<\delta N^2 \right> = L^{2(\chi+d)}  \times {\rm constant}
\label{delN8}
\end{eqnarray}
where the constant is independent of $L$.  This can once again  be rewritten in 
terms of the mean number $\left<N\right>$ of critters in the box, using
$L = \left({\left<N\right> \over\rho
_0 } \right)^{1 \over d}$.  This gives:
\begin{eqnarray}
\sqrt {<\delta N^2>} \sim <N>^{\phi(d)}
\label{Nfluc}
\end{eqnarray}
with
\begin{eqnarray}
\phi(d)=1+\frac{\chi}{d}   \ .
\label{phi4thcase}
\end{eqnarray}

Note that in {\it all} cases, we've just shown that the scaling of number  fluctuations with mean
number violates the ``law of large numbers'': the general rule that rms
number fluctuations scale like the square root of mean number.   The
fluctuations Eq.~(\ref{Nfluc}) are infinitely larger than this prediction in the limit of
mean number 
$\left<N\right> \to \infty$ for {\it all} spatial dimensions $d$; hence,
they are much larger than those found in most equilibrium 
and most non-equilibrium systems, since most of those obey the law of
large numbers.  In the next section we will show evidence from our simulations that such Giant number fluctuations do occur in $d=3$.

 \section{Simulations} \label{Simulations}

\subsection{The numerical model}
We test these predictions by using a slight modification of the Vicsek algorithm \cite{Vicsek} to incorporate vectorial noise in any number of dimensions \cite{gregoire2003moving}.
The algorithm is as follows: shift the particles in their direction of travel by a distance $v_0$. Now  compute an intermediate velocity $v^\prime$ based on an alignment interaction and an attractive/repulsive interaction with coefficient $b$. Normalize this, perturb it with Gaussian noise with magnitude $T/\sqrt{v_0}$, and normalize again.

That is, the new velocity ${\bf v_i}(t+1)$ of the $i$'th particle on the $t+1$ time step  can be expressed in terms of the velocities on the $t$ time step as:

\begin{equation}
{\bf v_i}(t)^\prime = \sum_{j\in r_{ij}<1}\left[ {\bf v_j}(t)+b (|r_{ij}|-r_0)\hat{r}_{ij}\right]
\label{Vicsek0}
\end{equation}

\begin{equation}
{\bf v_i}(t+1) =v_0\hat{N}\left[\hat{N}\left[{\bf v_i}(t)^\prime\right] + \frac{T}{\sqrt{v_0}} {\bf \Upsilon}_i (t)\right]
\label{Vicsek1}
\end{equation}
where we have used the notation $\hat{N}[{\bf x}]\equiv\hat{x}$ for any vector ${\bf x}$  and the annealed noise is given by a Gaussian random  vector variable ${\bf \Upsilon}_i$ with $\overline{ \Upsilon_{i\alpha}(t) \Upsilon_{j\beta}(t^\prime) } = \delta_{\alpha\beta}\delta_{ij} \delta(t-t^\prime)$, where $\alpha$ and $\beta$ denote Cartesian components. The factor of $\sqrt{v_0}$ is intended to correct for the effect that the shorter the movement step is, the more the noise self-averages out (just as one would integrate the effects of noise in a continuum Langevin equation using $\sqrt{\Delta t}$ instead of $\Delta t$).

Quenched disorder is implemented by adding a certain number of static particles (``dead birds") to the simulation. These particles are placed at fixed positions chosen randomly from a spatially uniform distribution, and assigned, also randomly,  fixed ``pseudo-velocity vectors" of length $v_0$,   with an isotropic distribution. These positions and pseudo-velocity vectors do not change throughout the simulation. Clearly, the ``pseudo-velocity vectors" of the dead birds do not correspond to their actual velocities (since those birds don't actually move), but they are treated like real velocity vectors in the evolution of the moving particles (the ``live" birds), albeit  with a weight $w$ that can be used to control the strength of the disorder. Dead birds also have $b=0$ and so do not interact through the repulsion term. To summarize, the first step of the two step algorithm to determine the new direction of motion, i.e., Eq.~(\ref{Vicsek0}), is replaced by:
\begin{equation}
{\bf v_i}(t)^\prime = \sum_{j\in r_{ij}<1} \left[ w_j {\bf v_j}(t)+b_j (|r_{ij}|-r_0)\hat{r}_{ij}\right] ~,
\label{Vicsek3}
\end{equation}
where $w_j=1$ for live birds and $w_j=w$ for dead birds, while $b_j=1$ for live birds and $b_j=0$
for dead birds.  In Eq.~(\ref{Vicsek3}), the sum is over {\it all} birds, {\it both} dead and alive, within a distance $1$ of the particular live bird whose velocity is being updated. The second step, and the motion step, are unaltered for the live birds, while the dead birds neither move, nor change the directions of their ``pseudo-velocity" vectors. When we compute system-wide averages, correlation functions, etc.,  we exclude these particles from the calculation.

We quantify the strength (variance) of the quenched disorder by defining a noise parameter $\Delta\equiv  w^2 \rho_D/\rho_0$, where $\rho_D$ and $\rho_0$ are the density of dead and live birds respectively. In general we consider systems with periodic boundaries of linear dimension $L$, with $r_0=0.9$, an interaction radius of $1$. We consider both cases in which there is only quenched disorder ($T=0$) and where there is both quenched and annealed disorder ($T\neq0$), as well as different values of $v_0$.

\subsection{Average Velocity and Velocity Correlations}

\begin{figure}
\includegraphics[trim={6.8cm 3.8cm 7.0cm 4cm}, clip=true,width=2.75in]{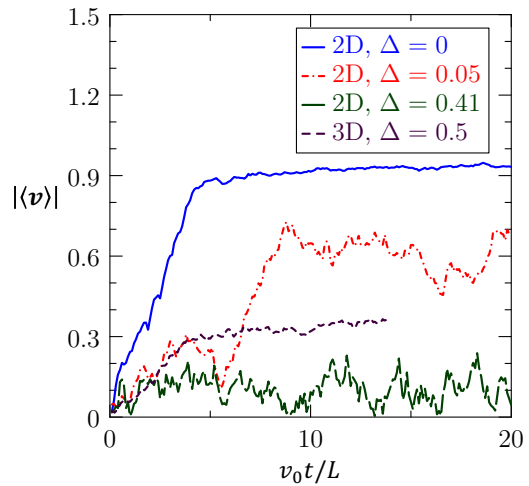}
 \caption{Convergence of $|\langle {\bf v}\rangle |$ for simulations of two and three dimensional flocks. The two dimensional simulations are at system size $L=100$ and the three dimensional simulations are at system size $L=128$. Data are plotted for various values of the disorder strength $\Delta$ versus time in units of the system transit time $L/v_0$.}
\label{FigConvergence}
\end{figure}

First we directly examine the order parameter $|\langle {\bf v}\rangle |$ where $\langle ... \rangle$ is the average over all particles in the system. If this quantity is non-zero, that indicates that the flock is in the ordered state. We simulate for enough time that particles could cross the system about $\ge 15$ times. In general we observe convergence to a plateau for about $8$ system transits (convergence plots of these data in two dimensions and three dimensions are in Fig.~\ref{FigConvergence}). Because of the system size, even if there is no true order in the infinite limit we expect to see the effect of finite size scaling in these data. These give rise to a residual order that should scale as $(L/L_0)^{-d/2}$ where $L_0$ is the lengthscale of patches of independent disorder (effectively corresponding to a Larkin length \cite{Larkin}).

\begin{figure}
 \includegraphics[trim={7.0cm 0 7.0cm  0}, clip=true,width=2.75in]{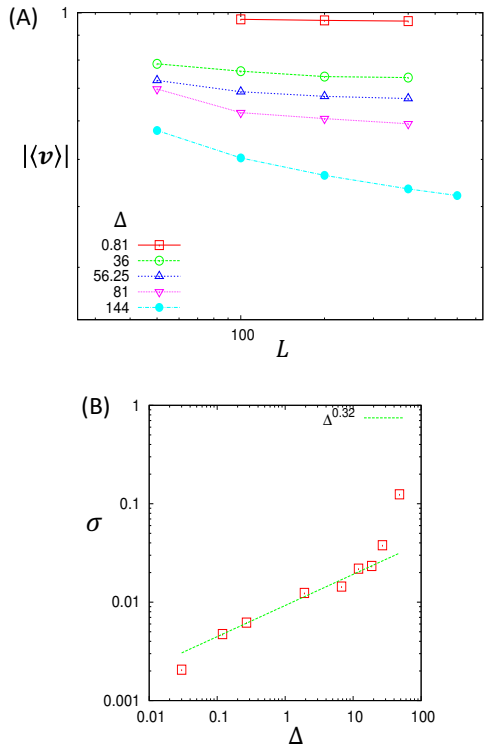}
\caption{Scaling of $|\langle v \rangle|$ with system size in two dimensions. (A) For large systems, the scaling is consistent with $L^{-\sigma/2}$ with a non-universal exponent $\sigma$ that depends on the disorder strength $\Delta$. (B) The exponent $\sigma$ increases with $\Delta$. In our model, it roughly follows a power law $\sigma \sim  \Delta ^{0.32}$ in a wide range of $\Delta$.}
\label{LScaling}
\end{figure}

We use systems of linear extents $L=64$ and $L=128$ in the three dimensional case, and systems of sizes $L=50$, $L=100$, $L=200$, and $L=400$ in the two dimensional case (in order to distinguish any true ordering from residual order originating from finite system size). For these data, we use a weight $w=1$ for the disorder particles. 
We find that in three dimensional systems there is long range order even for large disorder strength. For the system with $L=128$ and $\Delta=0.5$, the resulting value of the order parameter at long times is still $|\langle {\bf v} \rangle |=0.35$ (compared with $0.13$ for a comparable system in two
dimensions at $L=400$). In two dimensions, the average velocity $|\langle {\bf v} \rangle |$ decreases with system size $L$ albeit very slowly for small values of disorder $\Delta$ as shown in Fig.~\ref{LScaling}(A). The dependence of the average velocity on the system size $L$ can be fitted by a power law: $|\langle {\bf v} \rangle | \propto L^{-\sigma/2}$, where the non-universal exponent $\sigma$ increases with the disorder strength $\Delta$ as shown in Fig.~\ref{LScaling}(B).

\begin{figure}[!ht]
\includegraphics[trim={7.0cm 3cm 7.0cm 4cm}, clip=true,width=2.75in]{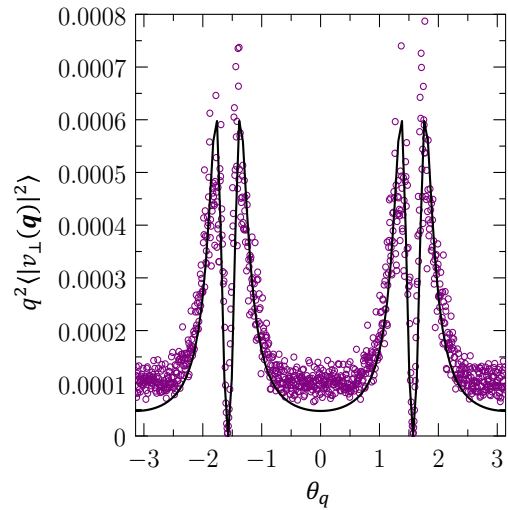}
\caption{Fourier-space velocity-velocity correlation function of a Vicsek swarm in the presence of quenched disorder (same as Fig. 1 in the short paper). The solid line is the theoretical prediction from the continuum hydrodynamic theory. The theory predicts a divergence at some critical angle $\theta_c$, and a zero at 90 degrees; in this case, $\theta_c \approx 78^\circ$.}
\label{Fig3Correlation}
\end{figure}

We have computed the Fourier transformed velocity-velocity correlation function in our simulations across multiple realizations of the disorder. The system size for these data is $L=600$, though we do not see a significant departure from these results in a simulations done at $L=1000$. We use parameters $v_0=0.1$, $T=0$, and $b=0$, with an average density $\rho_0=1$ and disorder strength $\Delta=0.3$. 
In  Fig.~\ref{Fig3Correlation}, we plot the simulation results for $q^2\langle |\bv(\bq) |^2\rangle $ versus the direction $\theta_q$ of $\bq$. The solid line in Fig.~\ref{Fig3Correlation} is from our prediction Eq.~(\ref{vq2}) for the two dimensional case with $\gvt>0$. As can be seen, the agreement is quite good. Note that the fit only has three parameters: the overall scale of  $q^2\langle |\bv(\bq) |^2\rangle $, the position$\theta_c$ of the peak, and the width $\delta$ of the peak. Note also that the fact that the data for different values of the magnitude $q$ of $\bq$ collapse onto a single curve when plotted versus $\theta_q$ is by itself decisive evidence for our prediction for the scaling of the full non-linear theory, and contradicts the predictions of the simple linear theory, which predicts that $q^2\langle |\bv(\bq) |^2 \rangle $ diverges like $1/q^2$ at $\theta_q=\theta_c$. The fact that there is a sharp, albeit finite peak at a particular propagation direction $\theta_c$ confirms that the model we simulated has $\gvt>0$.

\subsection{Density Correlation and Giant Number Fluctuations}

\begin{figure}[!h]
\includegraphics[trim={7.0cm 3cm 7.0cm 4cm}, clip=true,width=2.75in]{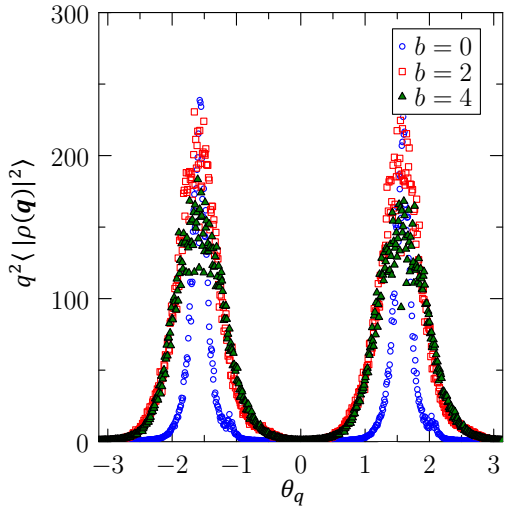}
\caption{Density autocorrelations at three different values of the repulsive interaction $b=0$, $b=2$, and $b=4$. Fits to the prediction of the hydrodynamic theory give values of $\theta_c$ of $89^\circ$, $86.8^\circ$, and $81.0^\circ$ respectively. }
\label{RhoCorrel}
\end{figure}

In Fig.~\ref{RhoCorrel}, we plot the Fourier transformed density-density correlation function
\begin{equation}
\langle| \rho({\bf q})|^2 \rangle= \frac{ \rho_0^2 \tan^2(\theta) \langle | v_\perp(\bf q )|^2\rangle}{v_2^2}
\end{equation}
versus $\theta_q$ for the same system for two dimensional system whose velocity correlations are plotted in (Fig.~\ref{Fig3Correlation}). We also show data for models identical to that of (Fig.~\ref{Fig3Correlation}) except that the value of the repulsion parameter $b$ has been increased to 
$b=2$ and $b=4$. Since increasing $b$ should increase $c_0$, which arises from pressure foces between the particles, we epect that $\theta_c=\tan^{-1}\sqrt{\gvt}/c_0$ should decrease as $b$ is increase. We indeed find that this is the case;
the values of $\theta_c$ for $b=0$, $b=2$, and $b=4$ are $89^\circ$, $86.8^\circ$, and $81.0^\circ$ respectively.

\begin{figure}[!h]
\includegraphics[width=2.75in]{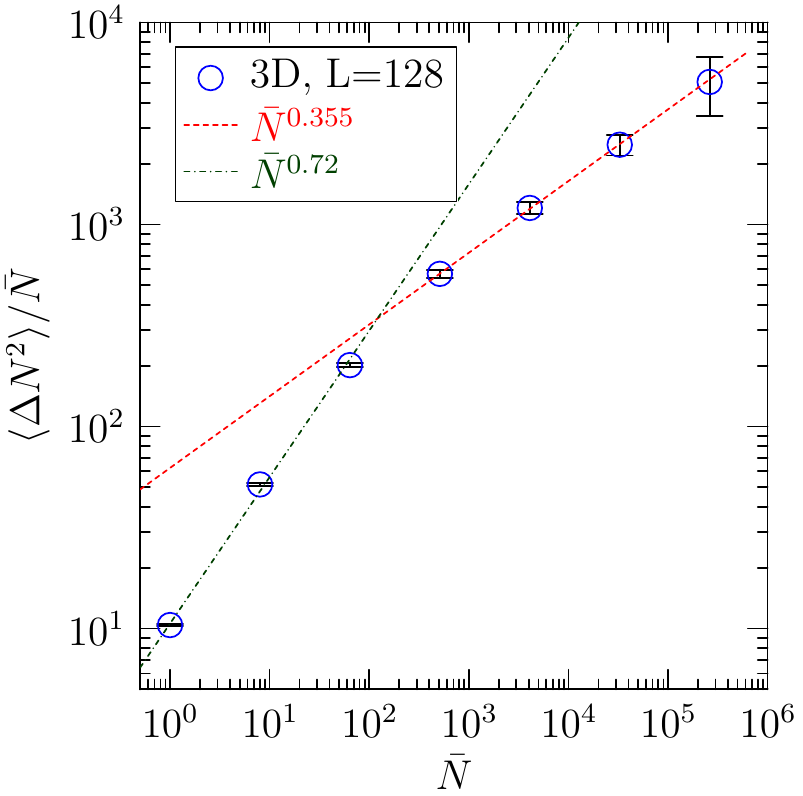}
\caption{Number fluctuations of a three dimensional Vicsek flock. The observed scalings correspond to $\phi=0.86$ (small $\bar{N}$) and $\phi=0.678$ (large $\bar{N}$).}
\label{GNFSFig}
\end{figure}

We also numerically determined the particle number fluctuations of a {\it  three} dimensional system of linear extent $L=128$ with quenced disorder strength $\Delta=0.5$ and annealed disorder strength $T=0.05$. We take the particle positions at the end of runs of duration $t=1700$ and decompose the system into boxes of linear length $2^i$ where $i=[0,6]$. These boxes are then used to measure the fluctuations $\langle \Delta N^2 \rangle(\bar{N})$. We average over five such simulations. The results are presented in Fig.~\ref{GNFSFig}. We observe two distinct regimes. At small length scales, we observe a scaling of $\langle \Delta N^2 \rangle / \bar{N} \propto \bar{N}^{0.72\pm0.03}$ corresponding to $\phi = 0.86\pm0.02$. This is presumably the small length scale behavior, before the non-linear effects become relevant. 
At larger scales, which we observe $\langle \Delta N^2 \rangle / \bar{N} \propto \bar{N}^{0.355\pm0.003}$, corresponding to $\phi = 0.678\pm0.002$. Comparing this with our prediction Eq.~(\ref{phi4thcase}) for the case $d=3$, $\gvt>0$, we see that this implies a somewhat surprisingly large negative value of $\chi=-.966$. More simulation studies in larger systems are needed to determine the exponents.

\section{Summary and Conclusions} 

We have developed a hydrodynamic theory of flocking in the presence of quenched disorder. The theory predicts that
flocks with non-zero quenched disorder can still develop long ranged order in three dimensions, and quasi-long-ranged order in two dimensions, in strong contrast to the equilibrium case, in which any amount of quenched disorder destroys ordering in both in two  and three dimensions \cite{Harris, Geoff, Aharonyrandom}. This prediction is consistent  with the results of Chepizhko et. al. \cite{Peruani}, who indeed find quasi-long-ranged order in $d=2$ systems with quenched disorder. 
We identify four qualitatively distinct cases, depending on the values of a combination of hydrodynamic parameters $\gvt$, and the dimension of space $d$. When $\gvt>0$, longitudinal sounds speeds in the flock vanish of certain critical angles $\theta_c$ between the
direction of propagation and \xpa, while for $\gvt<0$, those speeds are non-zero for all angles $\theta_q$  between the
direction of propagation and \xpa. Our hydrodynamic predicts that quenched disorder induces far larger fluctuations for wavevectors $\bq$ that lie along directions in which the longitudinal sounds speeds vanish, when such directions exist. Hence, flocks with $\gvt>0$ behave very differently from those with $\gvt<0$. 

There is also a profound difference between two dimensional systems ($d=2$) and systems in higher dimensions ($d>2$): the latter can have velocity fluctuations $\vt$ perpendicular to both \xpa and $\bq$, while the former cannot. When such velocity fluctuations 
{\it do} exist (i.e., in $d>2$, there is always directions of wavevector (specifically, $\theta_q=\pi/2$) for which fluctuations of $\vt$ are very large.

As a result, there are four distinct cases: A) $\gamma v_2 <0$, $d>2$; B) $\gamma v_2 <0$, $d=2$; C) $\gamma v_2 >0$, $d=2$; D) $\gamma v_2 >0$, $d>2$. We have developed both the linear and the non-linear theory for all four cases, and find exact scaling laws. with exact exponents, for fluctuations in the full non-linear theory for cases A, B, and C. We also find  scaling laws with unknown exponents for case D. We confirm many of these scaling laws with our  numerical simulations.

\section{Acknowledgements} We are very grateful to F. Peruani for invaluable discussions of his work in this subject, and to R. Das, M. Kumar, and S. Mishra for communicating their results to us prior to publication. JT thanks the Department of Physics, University of California, Berkeley, CA; the Aspen Center for Physics; the Kavli Institute for Theoretical Physics; the Institut f\"{u}r Theoretische Physik II: Weiche Materie,
Heinrich-Heine-Universit\"{a}t, D\"{u}sseldorf; the Max Planck Institute for the Physics of Complex Systems  Dresden; the Department of Bioengineering at Imperial College, London;  the Isaac Newton Institute, Cambridge, UK; The Higgs Centre for Theoretical Physics at the University of Edinburgh; and the Lorentz Center of Leiden University, for their hospitality while this work was underway.
He also thanks the  US NSF for support by
awards \#EF-1137815 and \#1006171; and the Simons Foundation for support by award \#225579. YT acknowledges support from NIH (R01-GM081747). 


\end{document}